\documentclass[aps,prd,groupedaddress, nofootinbib]{revtex4-1}
\usepackage{amsmath}
\usepackage{bbold}
\usepackage{dsfont}
\usepackage{rotating}
\usepackage{hyperref}
\usepackage{tikz}
\usetikzlibrary{fit}
\usetikzlibrary{backgrounds}
\usetikzlibrary{patterns}
\usepackage{pgfplots}
\usepackage{footmisc}
\usepackage{subfigure}
\begin{document}

\title{Redundancies in Explicitly Constructed Ten Dimensional Heterotic String Models}
\author{Timothy Renner}
\email{Timothy\_Renner@baylor.edu}
\author{Jared Greenwald}
\email{Jared\_Greenwald@baylor.edu}
\author{ Douglas Moore}
\email{Douglas\_Moore1@baylor.edu}
\author{Gerald Cleaver}
\email{Gerald\_Cleaver@baylor.edu}
\affiliation{Baylor University}
\date{\today}

\begin{abstract}
Using Baylor University's C++ software for construction of weakly coupled free fermionic heterotic string models, called the FF Framework, we explicitly construct the level 1 Ka$\check{c}$-Moody ten dimensional heterotic string models with the aim of understanding the redundancies endemic to this construction method. 
We show that for models in any even number of large space-time dimensions with a massless left mover and an odd ordered right mover, the maximal number of space-time supersymmetries are present. 
We show that in order to produce all of the models for a given order, different basis vectors must be built; one cannot vary only the GSO coefficients. 
We also show that all combinations of two order-2 basis vectors do not produce the same models as all possible single order-4 basis vectors, implying the product of the orders used in a search does not necessarily determine the models built. 
We also show that to build all of the D=10 level-1 models the inputs must be: sets of single order-6 basis vectors, pairs of basis vectors with orders 3 and 2, or sets of three order-2 basis vectors.  
\end{abstract}

\preprint{BU-HEPP-11-04}
\preprint{CASPER-11-03}

\maketitle

 \section{Introduction}\label{sec: Introduction} It was recently shown that possible string theory derived models which are physically consistent number at least approximately $10^{500}$. \cite{Bousso:2000, Ashok:2003} 
 The sheer volume of possible models has prompted computational and analytic examinations of the ``landscape" of possible string vacua.\cite{Donagi:2004, Valandro:2008, Balasubramanian:2008, Lebedev:2008, Gmeiner:2008, Dienes:2008, Gabella:2008, Donagi:2008} 
 The weakly coupled free fermionic heterotic string (WCFFHS) construction \cite{Antoniadis:1986, Antoniadis:1987, Kawai:1986_2} method has produced the most phenomenologically realistic models to date. \cite{Cleaver:1999, Lopez:1992, Faraggi:1989, Faraggi:1992, Antoniadis:1990, Leontaris:1999, Faraggi:1991, Faraggi:1992_2, Faraggi:1992_3, Faraggi:1991_2, Faraggi:1991_3, Faraggi:1995, Faraggi:1996, Cleaver:1997, Cleaver:1997_2, Cleaver:1997_3, Cleaver:1998, Cleaver:1998_2, Cleaver:1998_3, Cleaver:1998_4, Cleaver:1998_5, Cleaver:1999_2, Cleaver:1999_3, Cleaver:1999_4, Cleaver:2000, Cleaver:2000_2, Cleaver:2001, Cleaver:2001_2, Cleaver:2002, Cleaver:2002_2, Cleaver:2002_3, Perkins:2003, Perkins:2005, Cleaver:2008, Greenwald:2009, Cleaver:2011} 
 They are also somewhat easily produced via computer programs, and random basis vector and GSO coefficient sampling\cite{Dienes:2006, Dienes:2007_2} or statistical sampling of GSO coefficients for a fixed set of basis vectors\cite{Assel:2010} been performed. 
 However, random sample searches of the input parameters have several problems which are not trivial to address. \cite{Dienes:2007} 
 Systematic searches circumvent these problems by building all possible inputs given certain constraints.
 Algorithms for efficiently and systematically generating the input for WCFFHS models were reported in \cite{Robinson:2008}.
 Already such searches have revealed novel realizations of certain gauge groups. \cite{Obousy:2008}
 Baylor University's C++ software framework for this type of construction, called the FF Framework, is performing several systematic searches of models with ten and lower large space-time dimensions. 
 Since the ten dimensional models achievable in this manner are well known \cite{Kawai:1986} and the degrees of freedom for the inputs are fewest, the goal of this paper is to study the inputs producing the models and look at redundancies which occur. 
 These are then tested for models with fewer large space-time dimensions with the ultimate goal of reducing the total computation needed to perform searches in the space of four dimensional WCFFHS models, where realistic models appear.\\
 
The heterotic theories consist of closed strings which have independent sets of left and right moving modes. 
The left moving modes are ten dimensional superstrings, while the right moving modes are 26 dimensional bosonic strings. 
WCFFHS models are constructed by specifying the phases that fermion modes gain when parallel transported around non-contractible loops on a genus-1 world sheet of a string. 
The number of possible values these phases may take is referred to as the order, while the number of ``basis vectors" of phases used to specify the model is referred to as the layer. 
The layers form a basis set, and linear combinations of this basis set produce sectors from which the physical states of the model are built. 
The force and matter content is then determined from the symmetries of these states. \\
The following statements are true regarding this method:
 \begin{itemize}
 \item There are many redundancies endemic to this construction method.
 \item The number of unique models accessible to this method is finite.
 \end{itemize}
 These two statements imply that an understanding of the redundancies present in this construction at large space-time dimensions (D) above 4 could bring into reach the construction of all unique D=4 models by this method. 
 This is the motivation for the present study. 
 Additionally, further correlations between the basis vectors and the massless spectrum allow future searches to be more focused on producing semi-realistic models.

\section{D=10 Heterotic String Models in the Free Fermionic Construction}\label{sec: D=10_Heterotic_String_Models_in_the_Free_Fermionic_Construction}
The spectrum of D=10 heterotic string models has been well tabulated. \cite{Kawai:1986} 
All of these are presented in detail in \autoref{tab: D=10_Models}, except for an additional heterotic model containing a single $E_8$ at Ka$\check{c}$-Moody level 2. 
In the free fermionic construction, the latter model is constructed using real Ising fermions \footnote{Higher level Ka$\check{c}$-Moody algebras are possible with compact dimensions using left-right pairings, sometimes referred to as rank-cutting. 
For models in this study, all modes are paired as complex modes on the left and right moving parts of the basis vectors. 
Thus, higher level Ka$\check{c}$-Moody algebras will not appear in these data sets.}\cite{Kawai:1987, Lewellen:1989, Chaudhuri:1994, Aldazabal:1994, Cleaver:1995, Cleaver:1995_2, Cleaver:1996, Kakushadze:1996, Erler:1996, Kakushadze:1997} , which have not been included in the present study. 
\begin{table}
\begin{tabular}{||c|c||c|c||}
\hline \hline
Model&ST SUSY&Model&ST SUSY\\
\hline \hline 
$SO(32)$&
$1$&
$SO(32)$&
$0$\\
\hline
$\begin{array}{ccc}
E_8&\otimes &E_8
\end{array}$&
$1$&
$\begin{array}{ccc}
SO(16)&\otimes&SO(16)\\
\hline 
1&&128\\
128&&1\\
16&&16
\end{array}$&
$0$\\
\hline
$\begin{array}{ccc}
SO(8)&\otimes &SO(24)\\
\hline 
\overline{8}&&24\\
8&&24
\end{array}$&
$0$&
$\begin{array}{ccc}
SO(16)&\otimes&E_8\\
\hline 
\overline{128}&&1\\
128&&1
\end{array}$&
$0$\\
\hline
$\begin{array}{ccccccc}
SU(2)&\otimes &SU(2)&\otimes&E_7&\otimes&E_7\\
\hline
1&&2&&1&&56\\
1&&2&&56&&1\\
2&&1&&1&&56\\
2&&1&&56&&1
\end{array}$&
$0$&
$\begin{array}{ccc}
SU(16)&\otimes&U(1)\\
\hline
\overline{120}&&\\
\overline{120}&&\\
120&&\\
120&&
\end{array}$&
$0$\\
\hline
\end{tabular}
\caption{These are all possible D=10, level-1 models which can be constructed using the methods detailed. The dimensions of the non-Abelian matter representations are given underneath the respective gauge groups under which they transform. Abelian charges were not computed.}
\label{tab: D=10_Models}
\end{table}
The emphasis of the discussion to follow will be on the different manifestations of the construction parameters which produced the D=10, level-1 models. 
Such an examination will provide clues as to the redundancies inherent to the free fermionic heterotic construction, and will aid in more efficient systematic examinations of the input parameters for models with fewer large space-time dimensions.\\ 

All of the searches have the following parameters fixed:
\begin{itemize}
\item The first basis vector, $\vec{\mathbb{1}}$, in each model is completely made up of periodic modes $(\vec{1}^{~8}||\vec{1}^{~32})$, and serves as a canonical basis for the vectors. It is not shown when describing the input for a model.
\item The four left moving complex world-sheet fermions $\psi_c^{1},...,\psi^{4}_c$ (in light cone gauge) with space-time indices have all periodic boundary conditions and may contribute to the matter states in the model as well as the gauge states. The orders specified will be for the right moving part of the basis vector only, the left moving part will always be order 2. Thus, the total order remains $N$ for $N_{even}$, and $2N$ for $N_{odd}$. 
\item The number of ST SUSYs is found by counting gravitinos \footnote{This method does not give exact results for D=8 models.}.
\end{itemize}

\section{Searches With One Basis Vector} \label{sec: Searches_With_One_Basis_Vector}
\begin{figure}
\begin{tikzpicture}
\begin{axis}[xlabel=Order, ylabel=Unique Models]
\addplot[mark=x] coordinates{
(2,4) (3,2) (4,7) (5,2) (6,8) (7,2) (8,8) (9,2) (10,8) (11,2) (12,8) (13,2) (14,8)
};\end{axis}
\end{tikzpicture}
\caption{Plotted here are the number of unique models produced vs the order of the basis vectors which produced those models.}
\label{fig: L1_Unique_Models_vs_Order}
\end{figure}
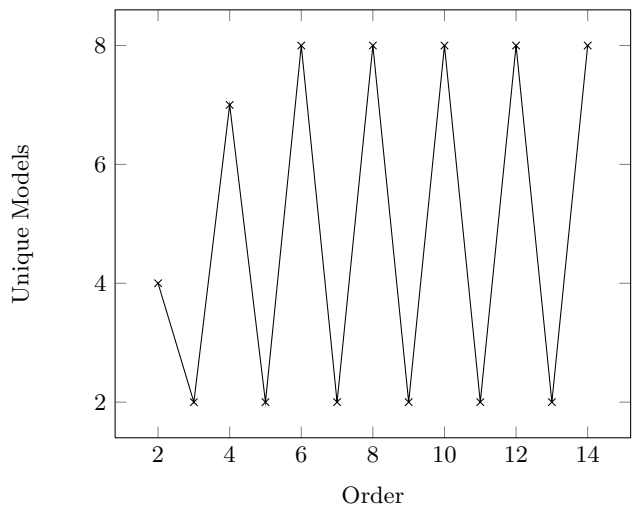

For the searches with a single basis vector (beyond the all-periodic), the order was increased with all possible basis vectors of that order investigated. \autoref{fig: L1_Unique_Models_vs_Order} shows the number of unique models vs the order of the basis vectors in the search. 
Notice that for odd orders, there are only two unique models constructed. For each of the odd orders, the models are $SO(32)$ and $E_8 \otimes E_8$, both with $N=1$ space-time supersymmetry. These are the only supersymmetric models present in the D=10 heterotic landscape. \\

ST SUSY in WCFFHS models, as mentioned in section \ref{sec: D=10_Heterotic_String_Models_in_the_Free_Fermionic_Construction}, is determined by counting gravitinos. 
Computationally, this involves first picking out all gravitino generating sectors as linear combinations of the basis vectors. 
This sector is identified as having a massless left moving part and all zeros for the right moving part. 
The only sector with this property is $(\vec{1}^{~8}||\vec{0}^{~32})$ (in ten large space-time dimensions). 
\footnote{The right moving space-time boson modes are customarily left out of WCFFHS models to save computing time, as the focus of WCFFHS phenomenology often involves analysis of the gauge and matter content, rather than gravity.}  
This sector is produced from every basis vector with odd order due to the coefficients which multiply it. 
For example, an order 3 basis vector has allowable phases of $0,~\frac{2}{3},~-\frac{2}{3}$ on the right side, and phases of $0,1$ on the left. 
Thus, the total order of the basis vector is $LCM(2,3)=6$. 
The coefficients which multiply that basis vector when constructing the sectors are 0 through 5. 
The right moving side has a $\mathds{Z}_3$ symmetry, which means that the values go through the following transformations:
\begin{equation}
\pm\left(\frac{2}{3}\right) \rightarrow^{\hspace{-1.1em}\times 0} 0 \rightarrow^{\hspace{-1.1em}\times 1} \pm\left(\frac{2}{3}\right) \rightarrow^{\hspace{-1.1em}\times 2} \mp\left(\frac{2}{3}\right)\rightarrow^{\hspace{-1.1em}\times3}0\rightarrow^{\hspace{-1.1em}\times4}\pm\left(\frac{2}{3}\right)\rightarrow^{\hspace{-1.1em}\times5}\mp\left(\frac{2}{3}\right)
\label{eqn: O3_RM_Sectors}
\end{equation}
The left moving side of the basis vector, however, has a $\mathds{Z}_2$ symmetry, and its values go though the following transformations:
\begin{equation}
1\rightarrow^{\hspace{-1.1em}\times0}0\rightarrow^{\hspace{-1.1em}\times1}1\rightarrow^{\hspace{-1.1em}\times2}0\rightarrow^{\hspace{-1.1em}\times3}1\rightarrow^{\hspace{-1.1em}\times4}0\rightarrow^{\hspace{-1.1em}\times5}1
\label{eqn: O3_LM_Sectors}
\end{equation}
Because the basis vector elements have the $\mathds{Z}_2^L || \mathds{Z}_3^R$ symmetry rather than a purely $\mathds{Z}_6$ symmetry, the gravitino generating sector always emerges. 
More generally, for any odd right moving order $N$, the basis vector (and consequently the associated sectors) has a $\mathds{Z}_2^L|| \mathds{Z}_{N}^R$ symmetry, rather than a purely $\mathds{Z}_{LCM(2,N)}$ symmetry. 
Therefore all basis vectors with odd right moving orders will produce a gravitino generating sector. 
We now conjecture that any model of even dimension with a single basis vector of odd order and massless left mover has the maximal number of space-time SUSYs. 
Models matching these conditions are presented in \autoref{tab: D6_D4_O3}. 
The table clearly shows this conjecture to be true for $D=4$ models and  $D=6$ models meeting these criteria. 
The conjecture was tested for order-5 models in $D=4$ and $D=6$ large space-time dimensions as well, with a sample of the relevant models (containing a massless left mover only) presented in \autoref{tab: D6_D4_O5}. 
All of those models (and the ones not shown explicitly) contain the maximal number of space-time supersymmetries.\\
\begin{table}
\begin{tabular}{||c|c|c|c|c||}
\hline
\hline
D&BV&$k_{ij}$&Model&N\\
\hline
\hline
6&$(\vec{1}^{~4}(1,0,0)^4||\vec{0}^{~34}(\vec{\frac{2}{3}})^6)$&$\left(\begin{smallmatrix}1&\\0&\end{smallmatrix}\right)$&
$\begin{array}{c}
SO(40)
\end{array}$&2\\
\hline
6&$(\vec{1}^{~4}(1,0,0)^4||\vec{0}^{~22}(\vec{\frac{2}{3}})^{18})$&$\left(\begin{smallmatrix}1&\\0&\end{smallmatrix}\right)$&
$\begin{array}{ccc}
SO(24)&\otimes&E_8
\end{array}$&2\\
\hline
6&$(\vec{1}^{~4}(1,0,0)^4||\vec{0}^{~16}(\vec{\frac{2}{3}})^{24})$&$\left(\begin{smallmatrix}1&\\0&\end{smallmatrix}\right)$&
$\begin{array}{ccc}
SO(16)&\otimes&SO(24)
\end{array}$&2\\
\hline
6&$(\vec{1}^{~4}(1,0,0)^4||\vec{0}^{~10}(\vec{\frac{2}{3}})^{30})$&$\left(\begin{smallmatrix}1&\\0&\end{smallmatrix}\right)$&
$\begin{array}{ccc}
SU(16)&\otimes&SO(10)
\end{array}$&2\\
\hline
6&$(\vec{1}^{~4}(1,0,0)^4||\vec{0}^{~4}(\vec{\frac{2}{3}})^{36})$&$\left(\begin{smallmatrix}1&\\0&\end{smallmatrix}\right)$&
$\begin{array}{ccc}
SU(2)^3&\otimes&SU(18)
\end{array}$&2\\
\hline
4&$(\vec{1}^{~2}(1,0,0)^6||\vec{0}^{~38}(\vec{\frac{2}{3}})^6)$&$\left(\begin{smallmatrix}1&\\0&\end{smallmatrix}\right)$&
$\begin{array}{c}
SO(44)
\end{array}$&4\\
\hline
4&$(\vec{1}^{~2}(1,0,0)^6||\vec{0}^{~26}(\vec{\frac{2}{3}})^{18})$&$\left(\begin{smallmatrix}1&\\0&\end{smallmatrix}\right)$&
$\begin{array}{ccc}
SO(28)&\otimes&E_8
\end{array}$&4\\
\hline
4&$(\vec{1}^{~2}(1,0,0)^6||\vec{0}^{~20}(\vec{\frac{2}{3}})^{24})$&$\left(\begin{smallmatrix}1&\\0&\end{smallmatrix}\right)$&
$\begin{array}{ccc}
SO(20)&\otimes&SO(24)
\end{array}$&4\\
\hline
4&$(\vec{1}^{~2}(1,0,0)^6||\vec{0}^{~12}(\vec{\frac{2}{3}})^{30})$&$\left(\begin{smallmatrix}1&\\0&\end{smallmatrix}\right)$&
$\begin{array}{ccc}
SU(16)&\otimes&SO(14)
\end{array}$&4\\
\hline
4&$(\vec{1}^{~2}(1,0,0)^6||\vec{0}^{~8}(\vec{\frac{2}{3}})^{36})$&$\left(\begin{smallmatrix}1&\\0&\end{smallmatrix}\right)$&
$\begin{array}{ccccc}
SU(2)&\otimes&SU(18)&\otimes&SO(8)
\end{array}$&4\\
\hline
4&$(\vec{1}^{~2}(1,0,0)^6||\vec{0}^{~2}(\vec{\frac{2}{3}})^{42})$&$\left(\begin{smallmatrix}1&\\0&\end{smallmatrix}\right)$&
$\begin{array}{ccccc}
SU(21)&\otimes&U(1)&\otimes&U(1)
\end{array}$&4\\
\hline
\hline
\end{tabular}
\caption{Order 3 models with six and four large space-time dimensions and massless left movers. This table provides evidence for a  conjecture that single basis vectors with odd order right movers always have the maximal number of space-time supersymmetries. Note also that only half of the $k_{ij}$ matrix is specified. The other half is constrained by modular invariance, and is therefore not a true degree of freedom for WCFFHS models. The basis vectors are presented in a real basis.}
\label{tab: D6_D4_O3}
\end{table}

\begin{table}
\begin{tabular}{||c|c|c|c|c||}
\hline
\hline
D&BV&$k_{ij}$&Model&N\\
\hline
\hline
6&$(\vec{1}^{~4}(1,0,0)^4|| \vec{0}^{~12}(\vec{\frac{2}{5}})^{14}(\vec{\frac{4}{5}})^{14})$&$\left(\begin{smallmatrix}1&\\0&\end{smallmatrix}\right)$&
$\begin{array}{ccccc}
SO(12)&\otimes&E_7&\otimes&E_7
\end{array}$&2\\
\hline
6&$(\vec{1}^{~4}(1,0,0)^4||\vec{0}^{~8}(\vec{\frac{2}{5}})^{16}(\vec{\frac{4}{5}})^{16})$&$\left(\begin{smallmatrix}1&\\0&\end{smallmatrix}\right)$&
$\begin{array}{ccccc}
SO(8)&\otimes&SO(16)&\otimes&SO(16)
\end{array}$&2\\
\hline
6&$(\vec{1}^{~4}(1,0,0)^4||\vec{0}^{~6}(\vec{\frac{2}{5}})^{32}(\vec{\frac{4}{5}})^2)$&$\left(\begin{smallmatrix}1&\\0&\end{smallmatrix}\right)$&
$\begin{array}{ccc}
SO(8)&\otimes&SO(32)
\end{array}$&2\\
\hline
6&$(\vec{1}^{~4}(1,0,0)^4||\vec{0}^{~6}(\vec{\frac{2}{5}})^{32}(\vec{\frac{4}{5}})^{12})$&$\left(\begin{smallmatrix}1&\\0&\end{smallmatrix}\right)$&
$\begin{array}{ccccc}
SU(4)&\otimes&SU(12)&\otimes&E_6
\end{array}$&2\\
\hline
6&$(\vec{1}^{~4}(1,0,0)^4||\vec{0}^{~4}(\vec{\frac{2}{5}})^{18}(\vec{\frac{4}{5}})^{18})$&$\left(\begin{smallmatrix}1&\\0&\end{smallmatrix}\right)$&
$\begin{array}{ccc}
SU(2)^2&\otimes&SO(10)^2
\end{array}$&2\\
\hline
6&$(\vec{1}^{~4}(1,0,0)^4||\vec{0}^{~2}(\vec{\frac{2}{5}})^{24}(\vec{\frac{4}{5}})^{14})$&$\left(\begin{smallmatrix}1&\\0&\end{smallmatrix}\right)$&
$\begin{array}{ccccc}
SU(12)&\otimes&SO(14)&\otimes&U(1)^2
\end{array}$&2\\
\hline
4&$(\vec{1}^{~2}(1,0,0)^6||\vec{0}^{~16}(\vec{\frac{2}{5}})^{14}(\vec{\frac{4}{5}})^{14})$&$\left(\begin{smallmatrix}1&\\0&\end{smallmatrix}\right)$&
$\begin{array}{ccccc}
SO(16)&\otimes&E_7&\otimes&E_7
\end{array}$&4\\
\hline
4&$(\vec{1}^{~2}(1,0,0)^{6}||\vec{0}^{~10}(\vec{\frac{2}{5}})^{22}(\vec{\frac{4}{5}})^{12})$&$\left(\begin{smallmatrix}1&\\0&\end{smallmatrix}\right)$&
$\begin{array}{ccccc}
SU(12)&\otimes&SO(10)&\otimes&E_6
\end{array}$&4\\
\hline
4&$(\vec{1}^{~2}(1,0,0)^6||\vec{0}^{~8}(\vec{\frac{2}{5}})^{18}(\vec{\frac{4}{5}})^{18})$&$\left(\begin{smallmatrix}1&\\0&\end{smallmatrix}\right)$&
$\begin{array}{ccccc}
SU(10)&\otimes&SU(10)&\otimes&SO(8)
\end{array}$&4\\
\hline
4&$(\vec{1}^{~2}(1,0,0)^6||\vec{0}^{~6}(\vec{\frac{2}{5}})^{24}(\vec{\frac{4}{5}})^{14})$&$\left(\begin{smallmatrix}1&\\0&\end{smallmatrix}\right)$&
$\begin{array}{ccccccc}
SU(4)&\otimes&SU(12)&\otimes&SO(14)&\otimes&U(1)
\end{array}$&4\\
\hline
4&$(\vec{1}^{~2}(1,0,0)^6||\vec{0}^{~4}(\vec{\frac{2}{5}})^{30}(\vec{\frac{4}{5}})^{10})$&$\left(\begin{smallmatrix}1&\\0&\end{smallmatrix}\right)$&
$\begin{array}{ccccc}
SU(2)^2&\otimes&SU(16)&\otimes&SO(10)
\end{array}$&4\\
\hline
4&$(\vec{1}^{~2}(1,0,0)^6||\vec{0}^{~4}(\vec{\frac{2}{5}})^{20}(\vec{\frac{4}{5}})^{20})$&$\left(\begin{smallmatrix}1&\\0&\end{smallmatrix}\right)$&
$\begin{array}{ccccc}
SU(2)^3&\otimes&SU(10)&\otimes&SU(10)
\end{array}$&4\\
\hline
\hline
\end{tabular}
\caption{A sample of order 5 models with six and four large space-time dimensions and massless left movers. All of them have the maximal number of space-time supersymmetries. The basis vectors are presented in a real basis.}
\label{tab: D6_D4_O5}
\end{table}
To prove this conjecture, one must consider how the gravitinos are created from the sector. Applying the raising and lowering operator, $\vec{F}$, according to the following equation generates possible gravitinos:
\begin{equation}
\vec{Q} = \frac{1}{2}\vec{\alpha} + \vec{F}\label{eqn: State}
\end{equation}
where $\vec{Q}$ is the state, $\vec{\alpha}$ is the sector which generated the state, and $\vec{F}$ is the raising/lowering operator. 
$\vec{F}$ is any vector consisting of $0,\pm 1$ such that the state is massless. 
\footnote{The conditions for masslessness are that the length squared of the left mover be equal to 1, while the length squared of the right mover be equal to 2.} 
The GSO projections will then choose which possible states generated by the $\vec{F}$'s in \autoref{eqn: State} is physically present in the model. 
The equation for the GSO projections is 
\begin{equation}
\vec{v}_j\cdot\vec{Q}_{\vec{\alpha}} = \sum_i a_ik_{ji} + s_j~~~~~(\text{mod 2})\label{eqn: GSOP}
\end{equation}
where $\vec{Q}_{\vec{\alpha}}$ is the state coming from the sector $\vec{\alpha}$, $a_i$ are the coefficients which produced the sector $\vec{\alpha}$, $k_{ji}$ are elements of the GSO coefficient matrix, and $s_i$ is equal to one if $\vec{v}_j$ is a space-time fermion sector, and 0 if $\vec{v}_j$ is a space-time boson sector. 
The dot product in equation \ref{eqn: GSOP} and the other equations in this section is a Lorentz dot product, where the dot products of the right movers are subtracted from the dot products of the left movers. 
Moreover, complex modes contribute twice as much to the dot product as real modes.\\

In the case of gravitinos, the mass shell condition for the right movers is ignored. 
For gravitinos, the space-time boson modes customarily left out of this construction method to save computing resources are raised. 
When the space-time boson modes are raised, none of the other right moving modes can be raised without giving the state mass. \\

The masslessness condition for the left movers is already fulfilled when the SUSY generating sector is produced. 
Since the raising operators will make the left mover massive and the lowering operators do not change the mass, only the lowering operators are applied. 
Lowering the space-time fermion modes will change the spin state of the same gravitino (within its given helicity). 
Only one gravitino helicity is allowed by the GSO projections per model in ten dimensions, while both can be present in dimensions lower than ten. 
Lowering the compactified modes (in models with less than ten large space-time dimensions) will create distinguishable gravitino states. 
This is the reason that D=6 models  have $N=2$ space-time SUSY, while D=4 models have $N=4$.\\

The possible gravitino states can be categorized by which compact fermion modes (and space-time modes for D=10) have been lowered with the $\vec{F}$ operator. 
These fall into one of two categories based on their dot products in the GSO projections \autoref{eqn: GSOP}. 
The possible gravitino states for D=10, 6, and 4 are listed in \autoref{tab: Possible_Gravitinos}. 
\begin{table}
\begin{tabular}{||c|c|c|c||}
\hline \hline
$D$&$\psi^{\mu}$&$x^i$&Dot\\
\hline
10& + + + +& &0\\
\hline
10&+ + + --& &1\\
\hline
\hline
6&+ + &+ +&0\\
\hline
6&+ +&+ --&1\\
\hline
6&+ +&-- +&1\\
\hline
6&+ +&-- --&0\\
\hline
\hline
4&+&+ + +&0\\
\hline
4&+&+ + --&1\\
\hline
4&+&+ -- +&1\\
\hline
4&+&-- + +&1\\
\hline
4&+&+ -- --&0\\
\hline
4&+&-- + --&0\\
\hline
4&+&-- -- +&0\\
\hline
4&+&-- -- --&1\\
\hline \hline
\end{tabular}
\caption{The possible gravitino states in 10, 6, and 4 large space-time dimensions. A + represents a charge value of $\frac{1}{2}$, while a -- represents a charge value of $-\frac{1}{2}$.  The dot products for both of the GSOP constraints are in this case identical. Note that the $y^i,~w^i$ values can also be periodic and thus can vary, but permutations of $x^i,~y^i,~w^i$ produce identical models when there is only one basis vector with a massless left mover. The states above are presented in a complex basis.}
\label{tab: Possible_Gravitinos}
\end{table}
The GSO projection equations will reveal why models with a single odd ordered basis vector and a massless left mover always have the maximal number of space-time SUSYs.
For possible gravitino states, \autoref{eqn: GSOP} can be simplified.
\begin{itemize}
\item The coefficients producing the sector for this class of model is always $(0,N_R)$ where $N_R$ is the order of the right mover.
\item $s_j$ is equal to 1 since the only basis vectors producing the model are $\vec{\mathbb{1}}$ and $\vec{v}$.
\end{itemize}

The conjecture can be shown heuristically by noting that with only two possible dot product values and two possible values for $k_{\vec{\mathds{1} \vec{v}}}$, the space of potential gravitino states can be divided into two parts. 
This gives a maximum space-time SUSY of N=1 for D=10, N=2 for D=6, and N=4 for D=4, which is known to be true. 
It can, with some effort, also be proven mathematically.

The GSOPs with the above conditions applied are 
\begin{eqnarray}
0 = N_R k_{\vec{\mathds{1}}\vec{v}} + 1\pmod{2}\\
0 = N_R k_{\vec{v} \vec{v}} + 1\pmod{2}\\
\end{eqnarray}
for ``even" gravitinos (dot product is equal to zero mod 2). For ``odd" gravitinos (dot product equal to one mod 2) the GSOPs are
\begin{eqnarray}
1 = N_R k_{\vec{\mathds{1}} \vec{v}} + 1\pmod{2}\\
1 = N_Rk_{\vec{v}\vec{v}} + 1\pmod{2}
\end{eqnarray}
This constrains the possible values that $k_{\vec{\mathds{1}}\vec{v}}$ and $k_{\vec{v}\vec{v}}$ can take in order for the possible gravitinos to survive the GSO projections.
\begin{eqnarray}
N_R k^e_{\vec{\mathds{1}} \vec{v}}=N_R k^e_{\vec{v} \vec{v}} = 1\pmod{2}\label{eqn: Even_Gravitino_Condition}\\
N_R k^o_{\vec{\mathds{1}} \vec{v}}=N_R k^o_{\vec{v} \vec{v}} = 0\pmod{2}\label{eqn: Odd_Gravitino_Condition}
\end{eqnarray}
where $k^e_{ij}$ is the GSO coefficient required for the even gravitino to pass, and $k^o_{ij}$ is the GSO coefficient required for odd gravitinos to pass. 
These are the conditions which must be proven to show that basis vectors with a massless left mover and odd ordered right mover always produce the maximal number of space-time supersymmetries.  
The conditions \autoref{eqn: Even_Gravitino_Condition}, and \autoref{eqn: Odd_Gravitino_Condition} imply that choosing $k_{\vec{v} \vec{\mathds{1}}}$ only ever eliminates half the total possible states, giving the model the maximal number of gravitinos. 

The proof will proceed as follows. The conditions \autoref{eqn: Even_Gravitino_Condition} and \autoref{eqn: Odd_Gravitino_Condition} will be rewritten in terms of the order-2 GSO coefficient $k_{\vec{v} \vec{\mathds{1}}}$ using the modular invariance constraints for the GSO coefficient matrix
\begin{eqnarray}
N_j k_{ij}=0\pmod{2}\label{eqn: k_ij_MI_1}\\
k_{ij} + k_{ji} = \frac{1}{2} \vec{v}_i \cdot \vec{v}_j\pmod{2}\label{eqn: k_ij_MI_2}\\
k_{ii} + k_{i1} = \frac{1}{4} \vec{v}_i \cdot \vec{v}_i - s_i\pmod{2}\label{eqn: k_ij_MI_3}
\end{eqnarray}
where $s_i$ is defined as in \autoref{eqn: GSOP}. 
This places a condition on the dot products of the basis vectors $\vec{\mathds{1}}$ and $\vec{v}$. 
A contradiction will be assumed and proven to be logically inconsistent, thus proving the following generalization of \autoref{eqn: Even_Gravitino_Condition} and \autoref{eqn: Odd_Gravitino_Condition}
\begin{equation}
N_Rk_{\vec{\mathds{1}} \vec{v}} = N_R k_{\vec{v} \vec{v}}\pmod{2}\label{eqn: General_Gravitino_Condition}
\end{equation}
which is sufficient to prove the conjecture.

Applying the modular invariance conditions to the above conditions results in
\begin{eqnarray}
N_R k_{\vec{\mathds{1}} \vec{v}} &=& -N_R k_{\vec{v} \vec{\mathds{1}}} - \frac{N_R}{2}\vec{\mathds{1}}\cdot \vec{v}\pmod{2}\\
N_R k_{\vec{v} \vec{v}} &=& -N_R k_{\vec{v}\vec{\mathds{1}}} - \frac{N_R}{4} \vec{v} \cdot \vec{v} - N_R\pmod{2}
\end{eqnarray}
Combining these conditions implies
\begin{equation}
-\frac{N_R}{2} \vec{\mathds{1}} \cdot \vec{v} = -\frac{N_R}{4} \vec{v} \cdot \vec{v} - N_R\pmod{2}\label{eqn: Gravitino_Dot_Product_Condition}
\end{equation}
Note that $N_R\pmod{2}$ is one. 
Modular invariance of basis vectors constrains the possible values for the dot products in the above equation. These constraints in general are
\begin{eqnarray}
N_{ij}\vec{v}_i\cdot \vec{v}_j&=&0\pmod{4}\\
N_{ii}\vec{v}_i\cdot \vec{v}_i&=&0\pmod{8}(\text{for even N})
\end{eqnarray}
where $N_{ij}$ is the least common multiple of the orders of the basis vectors $\vec{v}_{i,j}$. 
For the cases being considered, this implies
\begin{eqnarray}
2N_R \vec{\mathds{1}}\cdot \vec{v}&=&0\pmod{4}\implies\\
N_R \vec{\mathds{1}}\cdot \vec{v}&=&0\pmod{2}\\
2N_R\vec{v}\cdot \vec{v}&=&0\pmod{8}\implies \\
N_R\vec{v}\cdot \vec{v}&=&0\pmod{4}
\end{eqnarray}
Thus, the dot product terms in \autoref{eqn: Gravitino_Dot_Product_Condition} are integral. 
A contradiction can be used since either side can only be zero or one.

 Now the faulty assumption will be applied. 
 Assume, instead of \autoref{eqn: Gravitino_Dot_Product_Condition}, that the following is true
\begin{equation}
\frac{N_R}{2} \vec{\mathds{1}}\cdot \vec{v} = \frac{N_R}{4} \vec{v}\cdot \vec{v}\pmod{2}\label{eqn: False_Gravitino_Dot_Product_Condition}
\end{equation}
Splitting the Lorentz dot products into left and right movers, this condition can be further reduced by noting that $\vec{\mathds{1}}_L\cdot \vec{v}_L=\vec{v}_L\cdot \vec{v}_L = 4$. 
In terms of the right movers only, the equation becomes
\begin{equation}
N_R - \frac{N_R}{2} \vec{\mathds{1}}_R \cdot \vec{v}_R = - \frac{N_R}{4} \vec{v}_R\cdot \vec{v}_R\pmod{2}
\end{equation} 
where $N_R=1\pmod{2}$. 
The basis vector $\vec{v}$ can be split into its numerator and denominator.
\begin{eqnarray}
1-\frac{N_R}{2N_R} \vec{\mathds{1}}^N_R \cdot \vec{v}^N_R &=& -\frac{N_R}{4N_R^2} \vec{v}^N_R\cdot \vec{v}^N_R\pmod{2}\implies \nonumber \\
1-\frac{1}{2} \vec{\mathds{1}}^N_R\cdot \vec{v}^N_R &=& - \frac{1}{4N_R} \vec{v}^N_R\cdot \vec{v}^N_R\pmod{2}
\end{eqnarray}
The denominator of the right mover is $N_R$ due to the order of the right movers. 
Additionally, the numerators are all even integers. 
This comes from the constraint
\begin{equation}
N_R\vec{v}_R=0\pmod{2}
\end{equation}
The dot products can now be written as a summation.
\begin{eqnarray}
1-\frac{2}{2}\sum_i m_i &=& -\frac{4}{4N_R} \sum_i m_i^2\pmod{2}\nonumber \implies\\
1-\sum_im_i &=& \frac{1}{N_R} \sum_im_i^2\pmod{2}
\end{eqnarray}
where $m_i$ is an integer and the sum is over the indices of the right mover. 
Changing the basis to a multiplicative one, in which the value of the basis vector element is the index and the number of elements with that value, $N_i$ is summed over, we have
\begin{eqnarray}
1-\sum_{i=1}^{\lfloor N_R/2 \rfloor} iN_i &=& \frac{1}{N_R} \sum_{i=1}^{\lfloor N_R/2 \rfloor} i^2 N_i\pmod{2}\nonumber \implies\\
1&=&\frac{1}{N_R} \sum_{i=1}^{\lfloor N_R/2 \rfloor} i(i - N_R) N_i\pmod{2}\label{eqn: Final_Gravitino_Constraint}
\end{eqnarray}
Considering a case by case basis, assume $i$ is even. 
This means $i(i-N_R)N_i$ is even. 
If $i$ is odd, then $(i-N_R)$ is even, so $i(i-N_R)N_i$ is also even. 
Therefore every term in the sum is even, so the total sum is even. This implies that if the sum is a multiple of $N_R$, it is an even multiple of $N_R$, and the right hand side of \autoref{eqn: Final_Gravitino_Constraint} is $0\pmod{2}$ which is logically inconsistent since the left hand side is $1\pmod{2}$. 
If the sum is not a multiple of $N_R$, then it is a non-integral rational number which is also not equal to $1\pmod{2}$. 
Therefore \autoref{eqn: Final_Gravitino_Constraint} is impossible to satisfy, which show the faulty assumption \autoref{eqn: False_Gravitino_Dot_Product_Condition} is logically impossible. 
This means \autoref{eqn: Gravitino_Dot_Product_Condition} is true, and the conditions for each (odd or even) gravitino sectors are always satisfied. 
This proves the conjecture. \\

Moreover, it can be shown that one choice of $k_{\vec{v}\vec{\mathds{1}}}$ selects the even gravitinos, while the other choice selects the odd. 
Consider the modular invariance condition \autoref{eqn: k_ij_MI_2} applied to this scenario
\begin{equation}
k_{\vec{\mathds{1}} \vec{v}} + k_{\vec{v} \vec{\mathds{1}}} = \frac{1}{2}\vec{\mathds{1}}\cdot \vec{v}\pmod{2}
\end{equation}
 By \autoref{eqn: k_ij_MI_1}, $k_{\vec{v} \vec{\mathds{1}}}$ can only have a value of 0 or 1, and $k_{\vec{\mathds{1}}\vec{v}}$ is a rational number with denominator $N_R$. 
 Therefore evenness of the numerator determines whether $N_R k_{\vec{\mathds{1}}\vec{v}}~(\text{mod 2})$ is 0 or 1, passing either the odd or even gravitinos. 
 Using this information, it is clear that the two choices for $k_{\vec{v} \vec{\mathds{1}}}$ yield
 \begin{eqnarray}
 \frac{k_{\vec{\mathds{1}} \vec{v}}^N}{N_R}  = \frac{1}{2} \vec{\mathds{1}}\cdot \vec{v}\pmod{2}\\
 \frac{k_{\vec{\mathds{1}} \vec{v}}^N+N_R}{N_R}=\frac{1}{2} \vec{\mathds{1}}\cdot \vec{v}\pmod{2}
 \end{eqnarray} 
 where $k_{ij}^N$ is the numerator of the GSO coefficient. 
 Noting that the right hand side of each equation is an integer, combining them leads to
 \begin{equation}
 \frac{1}{N_R} (k_{\vec{\mathds{1}}\vec{v}}^{N0} + k_{\vec{\mathds{1}}\vec{v}}^{N1} + 1)=0\pmod{2}
 \end{equation}
 with the superscript indicating the choice of $k_{\vec{v} \vec{\mathds{1}}}$. 
 This equation is only satisfied if the quantity in parentheses is an even multiple of $N_R$. 
 This is true if and only if $k_{\vec{\mathds{1}} \vec{v}}^{N0} \neq k_{\vec{\mathds{1}}\vec{v}}^{N1}\pmod{2}$, which means
 \begin{equation}
 N_Rk_{\vec{\mathds{1}} \vec{v}}^{N0} \neq N_Rk_{\vec{\mathds{1}}\vec{v}}^{N1}\pmod{2}
 \end{equation}
 Ergo, choosing $k_{\vec{v} \vec{\mathds{1}}}$ admits either the even or the odd gravitinos into the model. \\
 
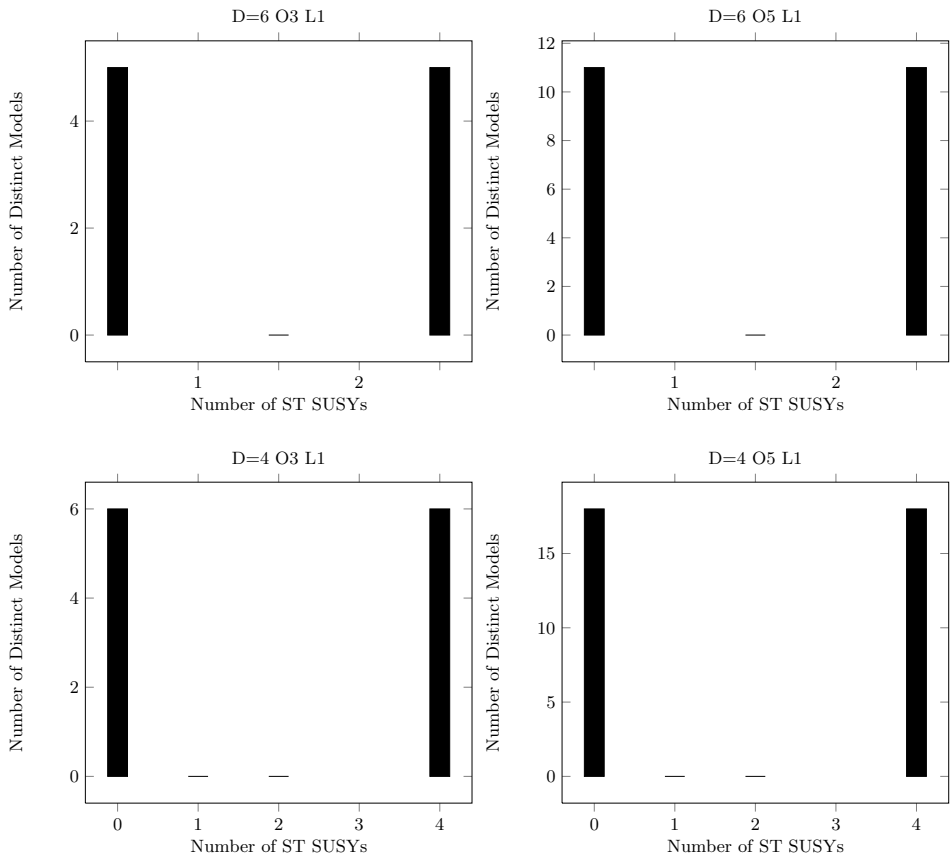
\begin{figure}
 \subfigure{
 \begin{tikzpicture}[scale=0.75]
\begin{axis} [ybar, ylabel = Number of Distinct Models, xlabel = Number of ST SUSYs, title={D=6 O3 L1}, xticklabels={0,,1,,2}]
\addplot[draw=black, fill=black]coordinates{
(0,5) (1, 0) (2,5) };
\end{axis}
\end{tikzpicture}
\begin{tikzpicture}[scale=0.75]
\begin{axis} [ybar, ylabel = Number of Distinct Models, xlabel = Number of ST SUSYs, title={D=6 O5 L1}, xticklabels={0,,1,,2}]
\addplot[draw=black, fill=black]coordinates{
(0,11) (1, 0) (2,11) };
\end{axis}
\end{tikzpicture}
}

 \subfigure{
 \begin{tikzpicture}[scale=0.75]
\begin{axis} [ybar, ylabel = Number of Distinct Models, xlabel = Number of ST SUSYs, title={D=4 O3 L1}]
\addplot[draw=black, fill=black]coordinates{
(0,6) (1, 0) (2, 0) (4,6) };
\end{axis}
\end{tikzpicture}
\begin{tikzpicture}[scale=0.75]
\begin{axis} [ybar, ylabel = Number of Distinct Models, xlabel = Number of ST SUSYs, title={D=4 O5 L1}]
\addplot[draw=black, fill=black]coordinates{
(0,18) (1, 0) (2, 0) (4,18) };
\end{axis}
\end{tikzpicture}
}
\caption{The number of distinct models against the number of space-time supersymmetries for D=6,4 O3,O5 L1. The number of distinct models with and without space-time SUSY are equal. The models themselves are also equal.}
\label{fig: D4D6_ST_SUSY}
 \end{figure}

 In addition to the conjecture above, there are also models in each of the four data sets with the exact same particle content without supersymmetry. 
 The basis vectors generating these models have the same right movers, but massive left movers \footnote{This does not happen for D=10, since there are not enough left moving modes to give the potential gravitino generating sector mass.}. 
 This removes the gravitino generating sector entirely from the model, leaving the model without supersymmetry. 
 This is shown in \autoref{fig: D4D6_ST_SUSY}, where the number of unique models is plotted against the number of space-time supersymmetries.
 The implications of the conclusions made in this section indicate that there is no correlation between the number of space-time supersymmetries and the gauge content for this class of models. More complicated models will need to be tested to determine whether such a correlation will emerge.
\section{Combinations with Two Order-2 Layers}\label{sec: Combinations_With_Two_Order_2_Layers}
\subsection{Varying $k_{ij}$'s} \label{sec: Varying_k_ij's}
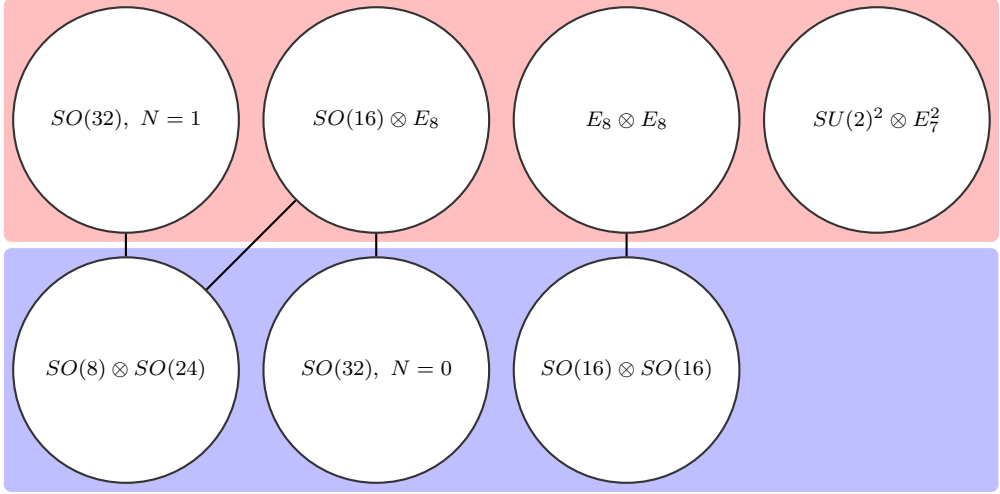
\begin{figure}
\tikzstyle{k_ij_1}=[rectangle,fill=red!25, rounded corners = 1mm]
\tikzstyle{k_ij_2}=[rectangle,fill=blue!25, rounded corners = 1mm]
\tikzstyle{Model_Input}=[rectangle, thick, draw=black!80, fill=white]
\tikzstyle{Model}=[circle, thick, minimum size=3cm, draw=black!80, fill=white]
\tikzstyle{Empty_Input}=[circle, minimum size=3cm]

\begin{tikzpicture}
\matrix[row sep = 0.3cm, column sep = 0.3cm]{
\node(Model_1_A)[Model]{\footnotesize{$SO(32),~N=1$}};&
\node(Model_2_A)[Model]{\footnotesize{$SO(16)\otimes E_8$}};&
\node(Model_3_A)[Model]{\footnotesize{$E_8\otimes E_8$}};&
\node(Model_4_A)[Model]{\footnotesize{$SU(2)^2 \otimes E_7^2$}};\\
\node(Model_1_B)[Model]{\footnotesize{$SO(8)\otimes SO(24)$}};&
\node(Model_2_B)[Model]{\footnotesize{$SO(32),~N=0$}};&
\node(Model_3_B)[Model]{\footnotesize{$SO(16)\otimes SO(16)$}};&
\node(Empty_4_B)[Empty_Input]{};\\
}
;
\path[-]
(Model_2_A) edge[thick] (Model_1_B)
(Model_1_A) edge[thick] (Model_1_B)
(Model_3_A) edge[thick] (Model_3_B)
(Model_2_A) edge[thick] (Model_2_B)
;

\begin{pgfonlayer}{background}
\node[k_ij_1, fit=(Model_1_A) (Model_4_A)]{};
\node[k_ij_2, fit=(Model_1_B) (Empty_4_B)]{};
\end{pgfonlayer}
\end{tikzpicture}
\caption{A schematic showing the systematic search for two basis vectors of order 2. The columns are models which are produced by different basis vectors, while the rows represent the possible $k_{ij}$ inputs. The lines indicate two models which were produced by the same basis vector set, but different $k_{ij}$ matrices. Therefore, a model with two lines was built by two different sets of basis vectors which produced different models when $k_{ij}$ was changed.}

\label{Fig: D10_O2_L1_O2_L2_k_ij}
\end{figure}

Figure \autoref{Fig: D10_O2_L1_O2_L2_k_ij} is a graphical representation of the basis vectors and the models they map to. It can be viewed as a ``summary" of the systematic search for this order/layer combination. It is important to note that different basis vectors must be used to build all of the models in the space; one cannot simply choose a basis set and only vary the GSO coefficient matrix (henceforth referred to as the $k_{ij}$ matrix) and get all of the models for these layer/order combinations. It would be worthwhile to find a set of basis vectors which produces all of the models by only varying the $k_{ij}$ matrix.\\

\subsection{Fixed $k_{ij}$'s} \label{sec: Fixed_k_ij's}
Fixing the $k_{ij}$ inputs to be equal to $\left(\begin{smallmatrix} 1&&\\1&&\\1&1&\end{smallmatrix}\right)$ does not prohibit solutions from being found - all of the models in \autoref{Fig: D10_O2_L1_O2_L2_k_ij} are present. 
The comparison of the search results between these two methods are presented in \autoref{tab: D10_O2_L1_O2_L2_Comparison}. 
Notice that if the $k_{ij}$ matrix entries for either search have only one value (either all 0's or all 1's) on the final row (the added basis vectors projecting onto the others in the GSOPs), then the basis vectors from both searches match. If the $k_{ij}$ matrix in the search on the left had multiple values for $k_{ij}$ matrix entries, then the corresponding fixed $k_{ij}$ basis vector has more ``breaks" - regions with matching boundary conditions. 
This suggests a relationship between the $k_{ij}$ matrix ``breaking" (meaning matching/non-matching values) and basis vector ``breaking."
\begin{table}
\begin{tabular}{||l|c||c||l|c||}
\hline
\multicolumn{2}{||c||}{O2 O2, varying $k_{ij}$}&Model&\multicolumn{2}{c||}{O2, O2, fixed $k_{ij}$}\\
\hline \hline
$\begin{array}{l}
(\vec{1}^{~8}||\vec{0}^{~32})\\
(\vec{1}^{~8}||\vec{0}^{~24}~\vec{1}^{~8})\\
\end{array}$ & 
$\left(\begin{smallmatrix}1&&\\0&&\\0&0&\end{smallmatrix}\right)$&
$SO(32),~N=1$&
$\begin{array}{l}
(\vec{1}^{~8} || \vec{0}^{~32})\\
(\vec{1}^{~8 }|| \vec{0}^{~24}\vec{1}^{~8})\\
\end{array}$&
$\left(\begin{smallmatrix} 1&&\\1&&\\1&1&\end{smallmatrix}\right)$\\
\hline
$\begin{array}{l}
(\vec{1}^8||\vec{0}^{32})\\
(\vec{1}^8||\vec{0}^{24}\vec{1}^8)\\
\end{array}$&
$\left(\begin{smallmatrix}1&&\\0&&\\0&1&\end{smallmatrix}\right)$&
$SO(8)\otimes SO(24)$&
$\begin{array}{l}
(\vec{1}^{~8}||\vec{0}^{~24}\vec{1}^{~8})\\
(\vec{1}^{~8}||\vec{0}^{~20}\vec{1}^{~4}\vec{0}^{~4}\vec{1}^{~4})\\
\end{array}$&
$\left(\begin{smallmatrix}1&&\\1&&\\1&1&\end{smallmatrix}\right)$\\
\hline
$\begin{array}{l}
(\vec{1}^{~8}||\vec{0}^{~32})\\[0.75ex]
(\vec{1}^{~8}||\vec{0}^{~16}\vec{1}^{~16})\\[1.5ex]
\end{array}$&
$\left(\begin{smallmatrix}1&&\\0&&\\0&0&\end{smallmatrix}\right)$&
$E_8\otimes E_8$&
$\begin{array}{l}
(\vec{1}^{~8}||\vec{0}^{~32})\\
(\vec{1}^{~8}||\vec{0}^{~16}\vec{1}^{~16})\\
\end{array}$&
$\left(\begin{smallmatrix}1&&\\1&&\\1&1&\end{smallmatrix}\right)$\\
\hline
$\begin{array}{l}
(\vec{1}^{~8}||\vec{0}^{~32})\\
(\vec{1}^{~8}||\vec{0}^{~16}\vec{1}^{~16})\\
\end{array}$&
$\left(\begin{smallmatrix}1&&\\0&&\\0&1& \end{smallmatrix}\right)$&
$SO(16)\otimes SO(16)$&
$\begin{array}{l}
(\vec{1}^{~8}||\vec{0}^{~24}\vec{1}^{~8})\\
(\vec{1}^{~8}||\vec{0}^{~16}\vec{1}^{~8}\vec{0}^{~8})\\
\end{array}$&
$\left(\begin{smallmatrix}1&&\\1&&\\1&1&\end{smallmatrix}\right)$\\
\hline
$\begin{array}{l}
(\vec{1}^{~8}||\vec{0}^{~24}\vec{1}^8)\\
(\vec{1}^{~8}||\vec{0}^{~16}\vec{1}^{~16})\\
\end{array}$&
$\left(\begin{smallmatrix}1&&\\0&&\\0&0&\end{smallmatrix}\right)$&
$SO(16)\otimes E_8$&
$\begin{array}{l}
(\vec{1}^{~8}||\vec{0}^{~24}\vec{1}^{~8})\\
(\vec{1}^{~8}||\vec{0}^{~16}\vec{1}^{~16})\\
\end{array}$&
$\left(\begin{smallmatrix}1&&\\1&&\\1&1&\end{smallmatrix}\right)$\\
\hline
$\begin{array}{l}
(\vec{1}^{~8}||\vec{0}^{~24}\vec{1}^{~8})\\
(\vec{1}^{~8}||\vec{0}^{~12}\vec{1}^{~12}\vec{0}^{~4}\vec{1}^{~4})\\
\end{array}$&
$\left(\begin{smallmatrix}1&&\\0&&\\0&0&\end{smallmatrix}\right)$&
$SU(2)^2\otimes E_7^2$&
$\begin{array}{l}
(\vec{1}^{~8}||\vec{0}^{~24}\vec{1}^{~8})\\
(\vec{1}^{~8}||\vec{0}^{~12}\vec{1}^{~12}\vec{0}^{~4}\vec{1}^{~4})\\
\end{array}$&
$\left(\begin{smallmatrix}1&&\\1&&\\1&1&\end{smallmatrix}\right)$\\
\hline
$\begin{array}{l}
(\vec{1}^{~8}||\vec{0}^{~16}\vec{1}^{~16})\\[0.75ex]
(\vec{1}^{~8}||\vec{0}^{~8}\vec{1}^{~24})\\[1.5ex]
\end{array}$&
$\left(\begin{smallmatrix}1&&\\0&&\\0&1&\end{smallmatrix}\right)$&
$SO(32),~N=0$&
$\begin{array}{l}
(\vec{1}^{~8}||\vec{0}^{~8}\vec{1}^{~24})\\[0.75ex]
(\vec{1}^{~8}||\vec{0}^{~4}\vec{1}^{~4}\vec{0}^{~4}\vec{1}^{~20})\\
\end{array}$&
$\left(\begin{smallmatrix}1&&\\1&&\\1&1&\end{smallmatrix}\right)$\\
\hline \hline
\end{tabular}
\caption{Comparison of the results between the searches in which the $k_{ij}$ matrix was and was not varied for two order 2 basis vectors. The inputs on the left were generated with multiple $k_{ij}$'s, while the inputs on the right were generated with a fixed $k_{ij}$.}
\label{tab: D10_O2_L1_O2_L2_Comparison}
\end{table}

\subsection{Relation to Order 4 Basis Vectors}\label{sec: Relation_to_Order_4_Basis_Vectors}
One may naively think that a single order-4 basis vector will be able to mimic the degrees of freedom available to two order-2 basis vectors, since $2\times 2=4$. 
However, closer examination makes it clear that the two data sets have different degrees of freedom. 
These degrees of freedom are best considered in the context of possible regions of matching boundary conditions, which determine the complexity of the model \footnote{In some cases there are enhancements to the gauge symmetries, so some higher layer/order models will actually produce fewer gauge groups of larger rank. These instances tend to be rare.}. 
Consider first a single basis vector of order-2. Since there are two values available for the basis vectors to acquire and, since all the boundary conditions for the all-periodic basis vector are identical, reordering of the elements (a redundancy for fermion modes which have matching boundary conditions in the other basis vectors in the model) gives two regions of matching boundary conditions. 
\begin{equation}
\begin{array}{c||cc}
(\vec{1}^{~8}&(0,...,0)&(1,...,1))
\end{array}
\end{equation}
Adding a second order 2 basis vector splits the regions up into four sets of matching boundary conditions.
\begin{equation}
\begin{array}{c||cccc}
(\vec{1}^{~8}&\multicolumn{2}{c}{(0,............,0)}&\multicolumn{2}{c}{(1,............,1)~~~~)}\\
(\vec{1}^{~8}&(0,...,0)&(1,...,1)&(0,...,0)&(1,...,1))
\end{array}
\end{equation}
For an order-4 basis vector, one assumes that there are still four regions of matching boundary conditions, $(0,\frac{1}{2},1,-\frac{1}{2})$. 
However, because the order-4 basis vector is the first layer with complex (neither integer nor half-integer) phases, there is a symmetry regarding the sign of the those phases. 
Effectively, 
\begin{equation}
-\frac{1}{2}\approx \frac{1}{2}
\end{equation}
for all complex phases in the basis vector. 
Thus, the regions of matching boundary conditions for a single order-4 basis vector are 
\begin{equation}
\begin{array}{c||ccc}
(\vec{1}^{~8}&(0,...,0)&(\frac{1}{2},...,\frac{1}{2})&(1,...,1))
\end{array}
\end{equation}
Additionally, the presence of complex phases add sectors to the model which produce fractional charge elements. 
Order-2 basis vectors produce only half-integer elements in the charge vectors coming from twisted sectors, while an order-4 basis vector produces not only the half integer twisted states, but also quarter integer twisted states. 
These additional degrees of freedom granted to the state vectors may result in additional symmetries which are unavailable to the sets of order-2 basis vectors.
Comparisons between these data sets have been tabulated in \autoref{tab: D10_O2_L1_O2_L2_O4_L1_Comparison}. 
\begin{table}
\begin{tabular}{||l|c||c||l|c||}
\hline \hline
\multicolumn{2}{||c||}{O2O2} & Model &\multicolumn{2}{c||}{O4}\\
\hline \hline
$\begin{array}{l}
(\vec{1}^{~8}||\vec{0}^{~32})\\
(\vec{1}^{~8}||\vec{0}^{~24}\vec{1}^{~8})
\end{array}$&
$\left(\begin{smallmatrix}1&&\\0&&\\0&0&\\\end{smallmatrix}\right)$&
$SO(32),~N=1$&
$\begin{array}{l}
(\vec{1}^{~8}||(\vec{\frac{2}{4}})^{32})
\end{array}$&
$\left(\begin{smallmatrix}1&\\0&\end{smallmatrix}\right)$\\
\hline
$\begin{array}{l}
(\vec{1}^{~8}||\vec{0}^{32})\\
(\vec{1}^{~8}||\vec{0}^{~24}\vec{1}^{~8})
\end{array}$&
$\left(\begin{smallmatrix}1&&\\0&&\\0&0&\end{smallmatrix}\right)$&
$SO(8)\otimes SO(24)$&
$\begin{array}{l}
(\vec{1}^{~8}||\vec{0}^{~22}(\vec{\frac{2}{4}})^{8}\vec{1}^{~2})
\end{array}$&
$\left(\begin{smallmatrix}1&\\0&\end{smallmatrix}\right)$\\
\hline
$\begin{array}{l}
(\vec{1}^{~8}||\vec{0}^{~32})\\
(\vec{1}^{~8}||\vec{0}^{~16}\vec{1}^{~16})
\end{array}$&
$\left(\begin{smallmatrix}1&&\\0&&\\0&0&\end{smallmatrix}\right)$&
$E_8\otimes E_8$&
N/A&
N/A\\
\hline
$\begin{array}{l}
(\vec{1}^{~8}||\vec{0}^{~32})\\
(\vec{1}^{~8}||\vec{0}^{~16}\vec{1}^{~16})
\end{array}$&
$\left(\begin{smallmatrix}1&&\\0&&\\0&1&\end{smallmatrix}\right)$&
$SO(16)\otimes SO(16)$&
$\begin{array}{l}
(\vec{1}^{~8}||\vec{0}^{~16}(\vec{\frac{2}{4}})^{16})
\end{array}$&
$\left(\begin{smallmatrix}1&\\0&\end{smallmatrix}\right)$\\
\hline
$\begin{array}{l}
(\vec{1}^{~8}||\vec{0}^{~24}\vec{1}^{~8})\\
(\vec{1}^{~8}||\vec{0}^{~16}\vec{1}^{~16})
\end{array}$&
$\left(\begin{smallmatrix}1&&\\0&&\\0&0&\end{smallmatrix}\right)$&
$SO(16)\otimes E_8$&
$\begin{array}{l}
(\vec{1}^{~8}||\vec{0}^{~14}(\vec{\frac{2}{4}})^{8}\vec{1}^{~10})
\end{array}$&
$\left(\begin{smallmatrix}1&\\0&\end{smallmatrix}\right)$\\
\hline
$\begin{array}{l}
(\vec{1}^{~8}||\vec{0}^{~24}\vec{1}^{~8})\\
(\vec{1}^{~8}||\vec{0}^{~12}\vec{1}^{~12}\vec{0}^{~4}\vec{1}^{~4})
\end{array}$&
$\left(\begin{smallmatrix}1&&\\0&&\\0&0&\end{smallmatrix}\right)$&
$SU(2)^2\otimes E_7^2$&
$\begin{array}{l}
(\vec{1}^{~8}||\vec{0}^{~12}(\vec{\frac{2}{4}})^{16}\vec{1}^{~4})
\end{array}$&
$\left(\begin{smallmatrix}1&\\0&\end{smallmatrix}\right)$\\
\hline
$\begin{array}{l}
(\vec{1}^{~8}||\vec{0}^{~16}\vec{1}^{~16})\\
(\vec{1}^{~8}||\vec{0}^{~8}\vec{1}^{~24})
\end{array}$&
$\left(\begin{smallmatrix}1&&\\0&&\\0&1&\end{smallmatrix}\right)$&
$SO(32),~N=0$&
$\begin{array}{l}
(\vec{1}^{~8}||\vec{0}^{~6}(\vec{\frac{2}{4}})^{8}\vec{1}^{~18})
\end{array}$&
$\left(\begin{smallmatrix}1&\\0&\end{smallmatrix}\right)$\\
\hline
N/A&
N/A&
$SU(16)\otimes U(1)$&
$\begin{array}{l}
(\vec{1}^{~8}||\vec{0}^{~6}(\vec{\frac{2}{4}})^{24}\vec{1}^{~2})
\end{array}$&
$\left(\begin{smallmatrix}1&\\0&\end{smallmatrix}\right)$\\[0.75ex]
\hline \hline
\end{tabular}
\caption{This table contains each model with the respective pair of order 2 and order 4 basis vectors, as well as the corresponding $k_{ij}$'s. Note that not all of the models can be produced from each data set. The basis vectors in this table are presented in a real basis.}
\label{tab: D10_O2_L1_O2_L2_O4_L1_Comparison}
\end{table}\\

 Note that the order 4 data set has the $SU(16)\otimes U(1)$ model, while it lacks the $E_8\otimes E_8$ model. 
 Untwisted boson sectors produce $SO(32)$ gauge groups in ten large space-time dimensions. 
 These are broken by the GSO projections of the twisted sectors into smaller $SO(2n)$ and $SU(n)$ groups. 
 To produce $SU(16)$, a rank 15 gauge group, the twisted sectors must span the entire group charge space. 
 In addition, the twisted sectors must be independent - that is, the basis vectors which produced them must be different - but not orthogonal. 
 This independence allows for contributions from one twisted sector to remove a root coming from the untwisted sector from the possible simple roots of the gauge group, making it an $SU(n)\otimes U(1)$ group rather than an $SO(2n)$ group. 
 A pair of order 2 basis vectors does not produce enough independent twisted sectors to do this. 
 The order 4 basis vector puts the weights of the adjoint representation in a twisted basis with charges of $\frac{1}{4}$, $-\frac{1}{4}$, $\frac{3}{4}$, $-\frac{3}{4}$ in addition to the half-integer charges, so three independent twisted sectors are not needed if there are more than half integer charges present in the model. \\
 
 By contrast, the $E_8\otimes E_8$ model does require the twisted sectors to be orthogonal and independent, which cannot be done with a single order-4 basis vector. 
 In an odd-ordered model, the lack of half integer twists produces orthogonal sectors in a twisted basis, allowing the $E_8\otimes E_8$ model to appear for those, but not for order-4 models.\\
 
 To summarize, the product of the orders does not determine the model spectrum - all possible combinations of orders must be investigated to fully map the model space produced by the free fermionic heterotic construction.

\section{The Full D=10, Level 1 Heterotic Landscape}\label{sec: The_Full_D=10_Landscape}
The full spectrum of D=10, level-1 models, as mentioned in \autoref{sec: D=10_Heterotic_String_Models_in_the_Free_Fermionic_Construction}, consists of the eight models presented in \autoref{tab: D=10_Models}. 
The lowest order for which all the models were built out of a single basis vector was order 6. 
It is worth testing whether one order-3 and one order-2 basis vector will also produce the full range of models, since $2\times 3 = 6$. 
While similar reasoning was not true for order-4 and two order-2's, this is indeed the case here. 
The models, and the basis vectors which produced them, are tabulated in \autoref{tab: D10_O6_O3O2}. 
Schematically, the O3O2 search is summarized by \autoref{fig: D10_O3_L1_O2_L2_k_ij}.
\begin{table}
\begin{tabular}{||l|c||c||l|c||}
\hline \hline
\multicolumn{2}{||c||}{O6}&Model&\multicolumn{2}{c||}{O3O2}\\
\hline \hline
$\begin{array}{l}
(\vec{1}^{~8}||\vec{0}^{~26}(\vec{\frac{2}{3}})^{6})
\end{array}$&
$\left(\begin{smallmatrix}1&\\0&\end{smallmatrix}\right)$&
$SO(32),~N=1$&
$\begin{array}{l}
(\vec{1}^{~8}||\vec{0}^{~26}(\vec{\frac{2}{3}})^{6})\\
(\vec{1}^{~8}||\vec{0}^{~24}\vec{1}^{~8})
\end{array}$&
$\left(\begin{smallmatrix}1&&\\1&&\\0&0&\end{smallmatrix}\right)$\\
\hline
$\begin{array}{l}
(\vec{1}^{~8}||\vec{0}^{~24}(\vec{\frac{1}{3}})^{6}\vec{1}^{~2})
\end{array}$&
$\left(\begin{smallmatrix}1&\\0&\end{smallmatrix}\right)$&
$SO(8)\otimes SO(24)$&
$\begin{array}{l}
(\vec{1}^{~8}||\vec{0}^{~26}(\vec{\frac{2}{3}})^{6})\\
(\vec{1}^{~8}||\vec{0}^{~24}\vec{1}^{~8})
\end{array}$&
$\left(\begin{smallmatrix}1&&\\0&&\\0&0&\end{smallmatrix}\right)$\\
\hline
$\begin{array}{l}
(\vec{1}^{~8}||\vec{0}^{~14}(\vec{\frac{1}{3}})^{12}\vec{1}^{~4})
\end{array}$&
$\left(\begin{smallmatrix}1&\\0&\end{smallmatrix}\right)$&
$SO(16)\otimes E_8$&
$\begin{array}{l}
(\vec{1}^{~8}|| \vec{0}^{~14}(\vec{\frac{2}{3}})^{18})\\
(\vec{1}^{~8}||\vec{0}^{~24}\vec{1}^{~8})
\end{array}$&
$\left(\begin{smallmatrix}1&&\\0&&\\0&\frac{2}{3}&\end{smallmatrix}\right)$\\
\hline
$\begin{array}{l}
(\vec{1}^{~8}||\vec{0}^{~14}(\vec{\frac{1}{3}})^{16}(\vec{\frac{2}{3}})^{2})
\end{array}$&
$\left(\begin{smallmatrix}1&\\0&\end{smallmatrix}\right)$&
$SO(16)\otimes SO(16)$&
$\begin{array}{l}
(\vec{1}^{~8}||\vec{0}^{26}(\vec{\frac{2}{3}})^{6})\\
(\vec{1}^{~8}||\vec{0}^{~16}\vec{1}^{~16})
\end{array}$&
$\left(\begin{smallmatrix}1&&\\0&&\\0&0&\end{smallmatrix}\right)$\\
\hline
$\begin{array}{l}
(\vec{1}^{~8}||\vec{0}^{~14}(\vec{\frac{2}{3}})^{18})
\end{array}$&
$\left(\begin{smallmatrix}1&\\0&\end{smallmatrix}\right)$&
$E_8\otimes E_8$&
$\begin{array}{l}
(\vec{1}^{~8}||\vec{0}^{~26}(\vec{\frac{2}{3}})^{6})\\
(\vec{1}^{~8}||\vec{0}^{~16}\vec{1}^{~16})
\end{array}$&
$\left(\begin{smallmatrix} 1&&\\1&&\\0&0&\end{smallmatrix}\right)$\\
\hline
$\begin{array}{l}
(\vec{1}^{~8}||\vec{0}^{~12}(\vec{\frac{1}{3}})^{14}(\vec{\frac{2}{3}})^4\vec{1}^{~2})
\end{array}$&
$\left(\begin{smallmatrix}1&\\0\end{smallmatrix}\right)$&
$SU(2)^2\otimes E_7^2$&
$\begin{array}{l}
(\vec{1}^{~8}||\vec{0}^{~14}(\vec{\frac{2}{3}})^{18})\\
(\vec{1}^{~8}||\vec{0}^{~12}\vec{1}^{~2}\vec{0}^{~12}\vec{1}^{~6})
\end{array}$&
$\left(\begin{smallmatrix}1&&\\0&&\\0&0&\end{smallmatrix}\right)$\\
\hline
$\begin{array}{l}
(\vec{1}^{~8}||(\vec{\frac{1}{3}})^{22}(\vec{\frac{2}{3}})^2\vec{1}^{~2})
\end{array}$&
$\left(\begin{smallmatrix}1&\\0&\end{smallmatrix}\right)$&
$SU(16)\otimes U(1)$&
$\begin{array}{l}
(\vec{1}^{~8}||\vec{0}^{~8}(\vec{\frac{2}{3}})^{~24})\\
(\vec{1}^{~8}||\vec{0}^{~6}\vec{1}^{~2}\vec{0}^{~18}\vec{1}^{~6})
\end{array}$&
$\left(\begin{smallmatrix}1&&\\1&&\\0&0&\end{smallmatrix}\right)$\\
\hline
$\begin{array}{l}
(\vec{1}^{~8}||\vec{0}^{~8}(\vec{\frac{1}{3}})^{~12}\vec{1}^{~12})
\end{array}$&
$\left(\begin{smallmatrix}1&\\0&\end{smallmatrix}\right)$&
$SO(32),~N=0$&
$\begin{array}{l}
(\vec{1}^{~8}||\vec{0}^{~8}(\vec{\frac{2}{3}})^{24})\\
(\vec{1}^{~8}||\vec{0}^{~8}\vec{1}^{~24})
\end{array}$&
$\left(\begin{smallmatrix}1&&\\1&&\\1&0&\end{smallmatrix}\right)$\\
\hline
\end{tabular}
\caption{Above are the models along with the inputs from each data set which produced that model. The basis vectors are presented in a real basis. Note that some of the order-6 basis vectors are actually order-3. Specifically, the $SO(32),~N=1$ model is produced by the same order-3 basis vectors. The additional order-2 basis vector's contribution is completely projected out.}
\label{tab: D10_O6_O3O2}
\end{table}
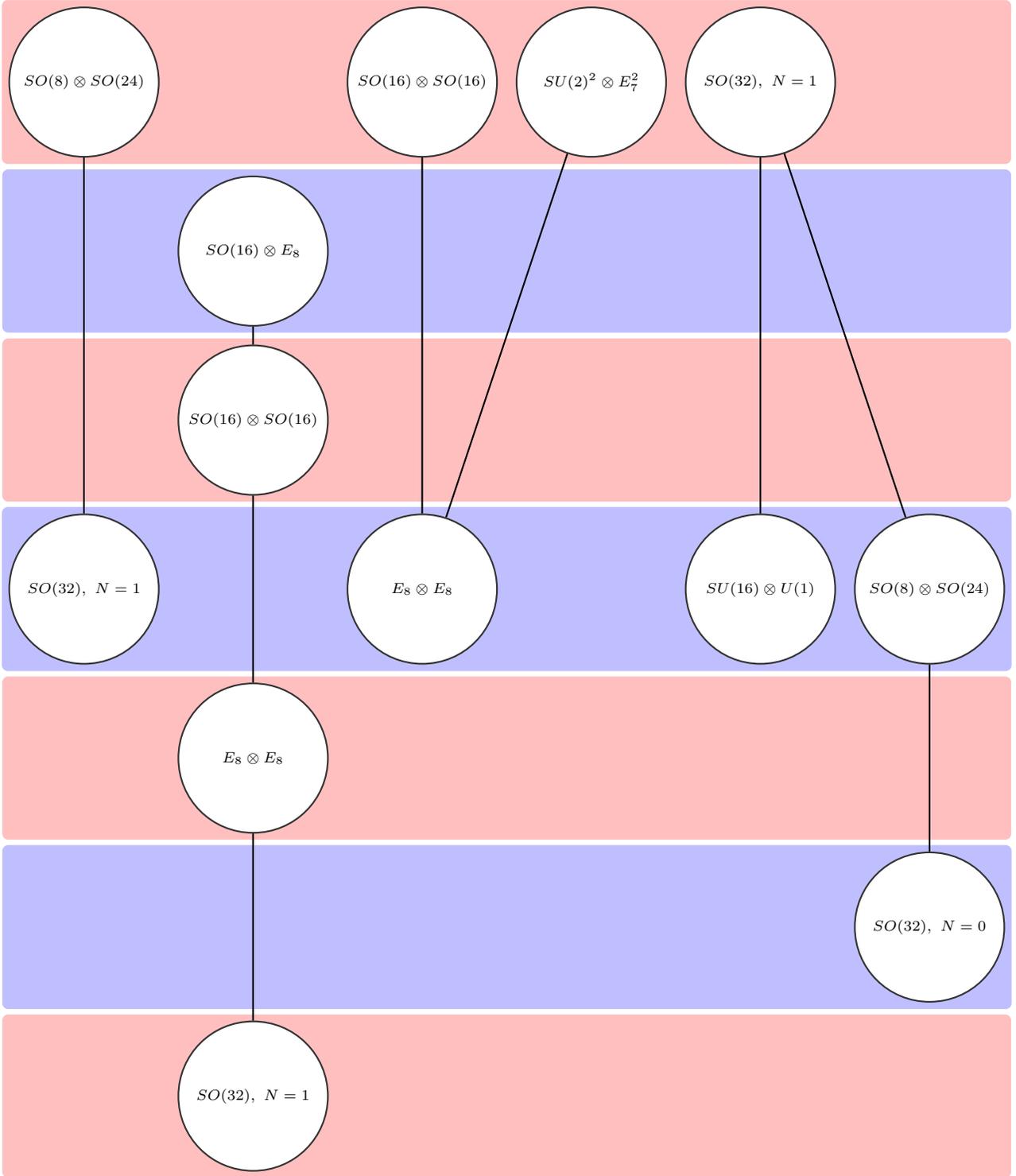
\begin{figure}
\tikzstyle{k_ij_1}=[rectangle,fill=red!25, rounded corners = 1mm]
\tikzstyle{k_ij_2}=[rectangle,fill=blue!25, rounded corners = 1mm]
\tikzstyle{Model_Input}=[rectangle, thick, draw=black!80, fill=white]
\tikzstyle{Model}=[circle, thick, minimum size=2.5cm, draw=black!80, fill=white]
\tikzstyle{Empty_Input}=[circle, minimum size=2.5cm]
\begin{tikzpicture} 
\matrix[row sep = 0.3cm, column sep = 0.3cm]{
\node(Model_1_A)[Model]{\scriptsize{$SO(8)\otimes SO(24)$}};&
\node(Empty_2_A)[Empty_Input]{};&
\node(Model_3_A)[Model]{\scriptsize{$SO(16)\otimes SO(16)$}};&
\node(Model_4_A)[Model]{\scriptsize{$SU(2)^2\otimes E_7^2$}};&
\node(Model_5_A)[Model]{\scriptsize{$SO(32),~N=1$}};&
\node(Empty_6_A)[Empty_Input]{};\\

\node(Empty_1_B)[Empty_Input]{};&
\node(Model_2_B)[Model]{\scriptsize{$SO(16)\otimes E_8$}};&
\node(Empty_3_B)[Empty_Input]{};&
\node(Empty_4_B)[Empty_Input]{};&
\node(Empty_5_B)[Empty_Input]{};&
\node(Empty_6_B)[Empty_Input]{};\\

\node(Empty_1_C)[Empty_Input]{};&
\node(Model_2_C)[Model]{\scriptsize{$SO(16)\otimes SO(16)$}};&
\node(Empty_3_C)[Empty_Input]{};&
\node(Empty_4_C)[Empty_Input]{};&
\node(Empty_5_C)[Empty_Input]{};&
\node(Empty_6_C)[Empty_Input]{};\\

\node(Model_1_D)[Model]{\scriptsize{$SO(32),~N=1$}};&
\node(Empty_2_D)[Empty_Input]{};&
\node(Model_3_D)[Model]{\scriptsize{$E_8\otimes E_8$}};&
\node(Empty_4_D)[Empty_Input]{};&
\node(Model_5_D)[Model]{\scriptsize{$SU(16)\otimes U(1)$}};&
\node(Model_6_D)[Model]{\scriptsize{$SO(8)\otimes SO(24)$}};\\

\node(Empty_1_E)[Empty_Input]{};&
\node(Model_2_E)[Model]{\scriptsize{$E_8\otimes E_8$}};&
\node(Empty_3_E)[Empty_Input]{};&
\node(Empty_4_E)[Empty_Input]{};&
\node(Empty_5_E)[Empty_Input]{};&
\node(Empty_6_E)[Empty_Input]{};\\

\node(Empty_1_F)[Empty_Input]{};&
\node(Empty_2_F)[Empty_Input]{};&
\node(Empty_3_F)[Empty_Input]{};&
\node(Empty_4_F)[Empty_Input]{};&
\node(Empty_5_F)[Empty_Input]{};&
\node(Model_6_F)[Model]{\scriptsize{$SO(32),~N=0$}};\\

\node(Empty_1_G)[Empty_Input]{};&
\node(Model_2_G)[Model]{\scriptsize{$SO(32),~N=1$}};&
\node(Empty_3_G)[Empty_Input]{};&
\node(Empty_4_G)[Empty_Input]{};&
\node(Empty_5_G)[Empty_Input]{};&
\node(Empty_6_G)[Empty_Input]{};\\
}
;

\path[-]
(Model_1_A) edge[thick] (Model_1_D)
(Model_2_B) edge[thick] (Model_2_C)
(Model_2_C) edge[thick] (Model_2_E)
(Model_2_E) edge[thick] (Model_2_G)
(Model_3_A) edge[thick] (Model_3_D)
(Model_4_A) edge[thick] (Model_3_D)
(Model_5_A) edge[thick] (Model_5_D)
(Model_5_A) edge[thick] (Model_6_D)
(Model_6_D) edge[thick] (Model_6_F)

;

\begin{pgfonlayer}{background}
\node[k_ij_1, fit=(Model_1_A) (Empty_6_A)]{};
\node[k_ij_2, fit=(Empty_1_B) (Empty_6_B)]{};
\node[k_ij_1, fit=(Empty_1_C) (Empty_6_C)]{};
\node[k_ij_2, fit=(Model_1_D) (Model_6_D)]{};
\node[k_ij_1, fit=(Empty_1_E) (Empty_6_E)]{};
\node[k_ij_2, fit=(Empty_1_F) (Model_6_F)]{};
\node[k_ij_1, fit=(Empty_1_G) (Empty_6_G)] {};

\end{pgfonlayer}
\end{tikzpicture}
\caption{A schematic diagram of the O3O2 systematic search. The different columns represent different basis vectors, while the different rows represent possible $k_{ij}$ matrix configurations. Lines connect models produced by the same basis vector, but different $k_{ij}$ matrices.}
\label{fig: D10_O3_L1_O2_L2_k_ij}
\end{figure}

Also of interest is whether the full D=10, level-1 model spectrum can be produced using only periodic/anti-periodic modes (order-2 basis vectors). 
It can indeed, but only with sets of three basis vectors. They are tabulated adjacent to their order-6 counterparts in \autoref{tab: D10_O6_O2O2O2}. 
Schematically, the search is summarized in \autoref{fig: D10_O2_L1_O2_L2_O2_L3_k_ij}.
\begin{table}
\begin{tabular}{||l|c||c||l|c||}
\hline \hline
\multicolumn{2}{||c||}{O6}&Model&\multicolumn{2}{c||}{O2O2O2}\\
\hline
$\begin{array}{l}
(\vec{1}^{~8}||\vec{0}^{~26}(\vec{\frac{2}{3}})^{6})
\end{array}$&
$\left(\begin{smallmatrix}1&\\0&\end{smallmatrix}\right)$&
$SO(32),~N=1$&
$\begin{array}{l}
(\vec{1}^{~8}||\vec{0}^{~32})\\
(\vec{1}^{~8}||\vec{0}^{~24}\vec{1}^{~8})\\
(\vec{1}^{~8}||\vec{0}^{~20}\vec{1}^{~4}\vec{0}^{~4}\vec{1}^{~4})
\end{array}$&
$\left(\begin{smallmatrix}1&&&\\0&&&\\0&0&&\\0&0&0&\end{smallmatrix}\right)$\\
\hline
$\begin{array}{l}
(\vec{1}^{~8}||\vec{0}^{~24}(\vec{\frac{1}{3}})^{6}\vec{1}^{~2})
\end{array}$&
$\left(\begin{smallmatrix} 1&\\0&\end{smallmatrix}\right)$&
$SO(8)\otimes SO(24)$&
$\begin{array}{l}
(\vec{1}^{~8}||\vec{0}^{~32})\\
(\vec{1}^{~8}||\vec{0}^{~24}\vec{1}^{~8})\\
(\vec{1}^{~8}||\vec{0}^{~20}\vec{1}^{~4}\vec{0}^{~4}\vec{1}^{~4})
\end{array}$&
$\left(\begin{smallmatrix}1&&&\\0&&&\\0&0&&\\0&1&0&\end{smallmatrix}\right)$\\
\hline
$\begin{array}{l}
(\vec{1}^{~8}||\vec{0}^{~16}(\vec{\frac{1}{3}})^{12}\vec{1}^{~4})
\end{array}$&
$\left(\begin{smallmatrix}1&\\0&\end{smallmatrix}\right)$&
$SO(16)\otimes E_8$&
$\begin{array}{l}
(\vec{1}^{~8}||\vec{0}^{~32})\\
(\vec{1}^{~8}||\vec{0}^{~24}\vec{1}^{~8})\\
(\vec{1}^{~8}||\vec{0}^{~16}\vec{1}^{~8}\vec{0}^{~8})
\end{array}$&
$\left(\begin{smallmatrix}1&&&\\0&&&\\0&1&&\\0&1&0&\end{smallmatrix}\right)$\\
\hline
$\begin{array}{l}
(\vec{1}^{~8}||\vec{0}^{~14}(\vec{\frac{1}{3}})^{16}(\vec{\frac{2}{3}})^{2})
\end{array}$&
$\left(\begin{smallmatrix}1&\\0&\end{smallmatrix}\right)$&
$SO(16)\otimes SO(16)$&
$\begin{array}{l}
(\vec{1}^{~8}||\vec{0}^{~32})\\
(\vec{1}^{~8}||\vec{0}^{~24}\vec{1}^{~8})\\
(\vec{1}^{~8}||\vec{0}^{~16}\vec{1}^{~8}\vec{0}^{~8})
\end{array}$&
$\left(\begin{smallmatrix}1&&&\\0&&&\\0&0&&\\0&1&0&\end{smallmatrix}\right)$\\
\hline
$\begin{array}{l}
(\vec{1}^{~8}||\vec{0}^{~14}(\vec{\frac{2}{3}})^{18})
\end{array}$&
$\left(\begin{smallmatrix}1&\\0&\end{smallmatrix}\right)$&
$E_8\otimes E_8$&
$\begin{array}{l}
(\vec{1}^{~8}||\vec{0}^{~32})\\
(\vec{1}^{~8}||\vec{0}^{~24}\vec{1}^{~8})\\
(\vec{1}^{~8}||\vec{0}^{~16}\vec{1}^{~8}\vec{0}^{~8})
\end{array}$&
$\left(\begin{smallmatrix}1&&&\\0&&&\\0&0&&\\0&0&1&\end{smallmatrix}\right)$\\
\hline
$\begin{array}{l}
(\vec{1}^{~8}||\vec{0}^{~12}(\vec{\frac{1}{3}})^{14}(\vec{\frac{2}{3}})^{4}\vec{1}^{~2})
\end{array}$&
$\left(\begin{smallmatrix}1&\\0&\end{smallmatrix}\right)$&
$SU(2)^2\otimes E_7^2$&
$\begin{array}{l}
(\vec{1}^{~8}||\vec{0}^{~32})\\
(\vec{1}^{~8}||\vec{0}^{~24}\vec{1}^{~8})\\
(\vec{1}^{~8}||\vec{0}^{~12}\vec{1}^{~12}\vec{0}^{~4}\vec{1}^{~4})
\end{array}$&
$\left(\begin{smallmatrix}1&&&\\0&&&\\0&1&&\\0&0&0&\end{smallmatrix}\right)$\\
\hline
$\begin{array}{l}
(\vec{1}^{~8}||\vec{0}^{~6}(\vec{\frac{1}{3}})^{22}(\vec{\frac{2}{3}})^{2}\vec{1}^{~2})
\end{array}$&
$\left(\begin{smallmatrix}1&\\0&\end{smallmatrix}\right)$&
$SU(16)\otimes U(1)$&
$\begin{array}{l}
(\vec{1}^{~8}||\vec{0}^{~24}\vec{1}^{~8})\\
(\vec{1}^{~8}||\vec{0}^{~12}\vec{1}^{~12}\vec{0}^{~4}\vec{1}^{~4})\\
(\vec{1}^{~8}||\vec{0}^{~6}\vec{1}^{~6}\vec{0}^{~6}\vec{1}^{~6}\vec{0}^{~2}\vec{1}^{~2}\vec{0}^{~2}\vec{1}^{~2})
\end{array}$&
$\left(\begin{smallmatrix} 1&&&\\0&&&\\0&0&&\\0&0&0&\end{smallmatrix}\right)$\\
\hline
$\begin{array}{l}
(\vec{1}^{~8}||\vec{0}^{~8}(\vec{\frac{1}{3}})^{12}\vec{1}^{~12})
\end{array}$&
$\left(\begin{smallmatrix}1&\\0&\end{smallmatrix}\right)$&
$SO(32),~N=0$&
$\begin{array}{l}
(\vec{1}^{~8}||\vec{0}^{~24}\vec{1}^{~8})\\
(\vec{1}^{~8}||\vec{0}^{~16}\vec{1}^{~16})\\
(\vec{1}^{~8}||\vec{0}^{~8}\vec{1}^{~8}\vec{0}^{~8}\vec{1}^{~8})
\end{array}$&
$\left(\begin{smallmatrix}1&&&\\0&&&\\0&0&&\\0&0&1&\end{smallmatrix}\right)$\\
\hline \hline
\end{tabular}
\caption{Above are the models along with the inputs from each data set which produced that model. The basis vectors are expressed in a real basis.}
\label{tab: D10_O6_O2O2O2}
\end{table}
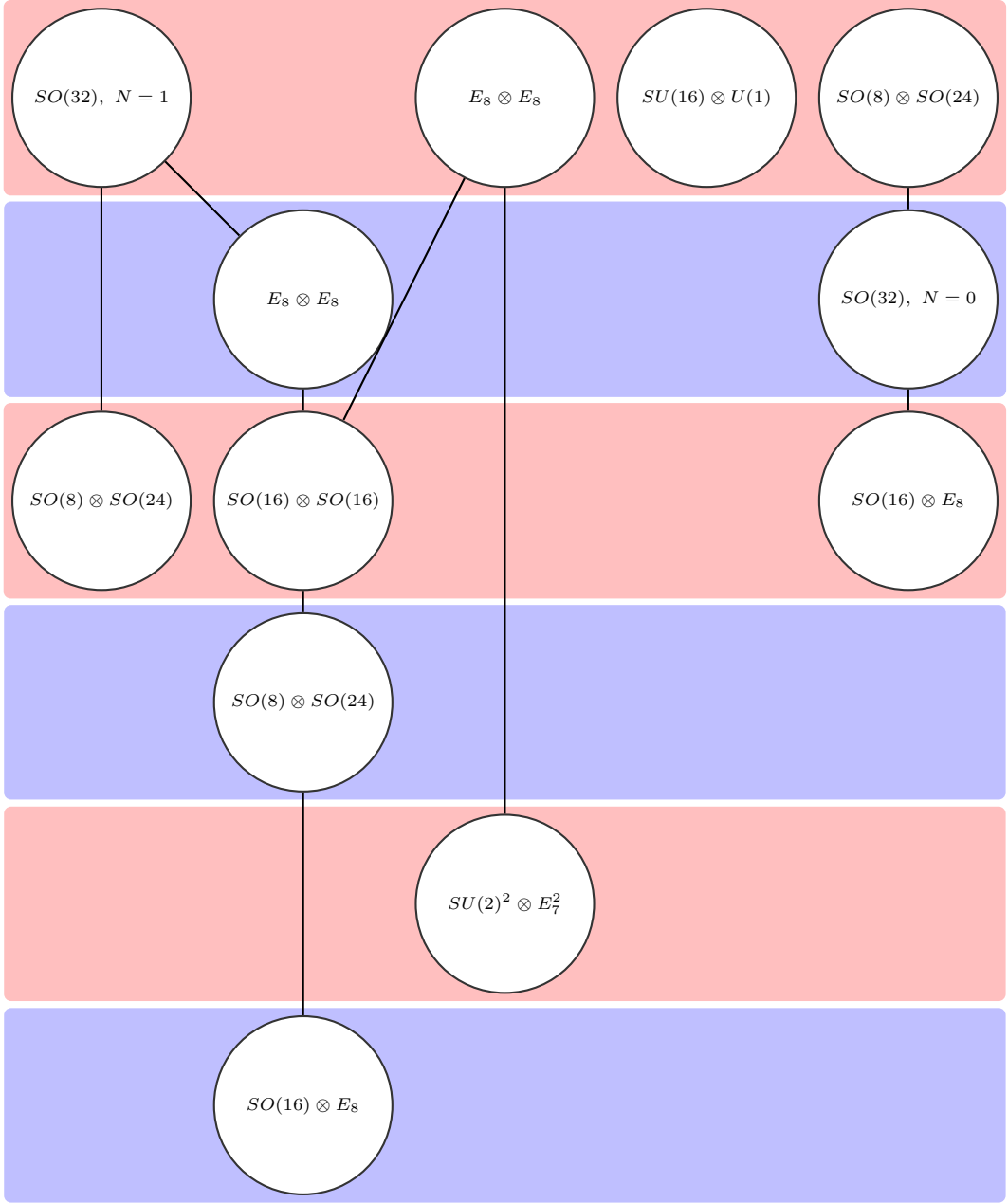
\begin{figure}
\tikzstyle{k_ij_1}=[rectangle,fill=red!25, rounded corners = 1mm]
\tikzstyle{k_ij_2}=[rectangle,fill=blue!25, rounded corners = 1mm]
\tikzstyle{Model_Input}=[rectangle, thick, draw=black!80, fill=white]
\tikzstyle{Model}=[circle, thick, minimum size=2.5cm, draw=black!80, fill=white]
\tikzstyle{Empty_Input}=[circle, minimum size=2.5cm]
\begin{tikzpicture} 
\matrix[row sep = 0.3cm, column sep = 0.3cm]{
\node(Model_1_A)[Model]{\scriptsize{$SO(32),~N=1$}};&
\node(Empty_2_A)[Empty_Input]{};&
\node(Model_3_A)[Model]{\scriptsize{$E_8\otimes E_8$}};&
\node(Model_4_A)[Model]{\scriptsize{$SU(16)\otimes U(1)$}};&
\node(Model_5_A)[Model]{\scriptsize{$SO(8)\otimes SO(24)$}};\\

\node(Empty_1_B)[Empty_Input]{};&
\node(Model_2_B)[Model]{\scriptsize{$E_8\otimes E_8$}};&
\node(Empty_3_B)[Empty_Input]{};&
\node(Empty_4_B)[Empty_Input]{};&
\node(Model_5_B)[Model]{\scriptsize{$SO(32),~N=0$}};\\

\node(Model_1_C)[Model]{\scriptsize{$SO(8)\otimes SO(24)$}};&
\node(Model_2_C)[Model]{\scriptsize{$SO(16)\otimes SO(16)$}};&
\node(Empty_3_C)[Empty_Input]{};&
\node(Empty_4_C)[Empty_Input]{};&
\node(Model_5_C)[Model]{\scriptsize{$SO(16)\otimes E_8$}};\\

\node(Empty_1_D)[Empty_Input]{};&
\node(Model_2_D)[Model]{\scriptsize{$SO(8)\otimes SO(24)$}};&
\node(Empty_3_D)[Empty_Input]{};&
\node(Empty_4_D)[Empty_Input]{};&
\node(Empty_5_D)[Empty_Input]{};\\

\node(Empty_1_E)[Empty_Input]{};&
\node(Empty_2_E)[Empty_Input]{};&
\node(Model_3_E)[Model]{\scriptsize{$SU(2)^2\otimes E_7^2$}};&
\node(Empty_4_E)[Empty_Input]{};&
\node(Empty_5_E)[Empty_Input]{};\\

\node(Empty_1_F)[Empty_Input]{};&
\node(Model_2_F)[Model]{\scriptsize{$SO(16)\otimes E_8$}};&
\node(Empty_3_F)[Empty_Input]{};&
\node(Empty_4_F)[Empty_Input]{};&
\node(Empty_5_F)[Empty_Input]{};\\
}
;

\path[-]
(Model_1_A) edge[thick] (Model_1_C)
(Model_1_A) edge[thick] (Model_2_B)
(Model_2_B) edge[thick] (Model_2_C)
(Model_2_C) edge[thick] (Model_2_D)
(Model_2_D) edge[thick] (Model_2_F)
(Model_3_A) edge[thick] (Model_2_C)
(Model_3_A) edge[thick] (Model_3_E)
(Model_5_A) edge[thick] (Model_5_B)
(Model_5_B) edge[thick] (Model_5_C)
;

\begin{pgfonlayer}{background}
\node[k_ij_1, fit=(Model_1_A) (Model_5_A)]{};
\node[k_ij_2, fit=(Empty_1_B) (Model_5_B)]{};
\node[k_ij_1, fit=(Model_1_C) (Model_5_C)]{};
\node[k_ij_2, fit=(Empty_1_D) (Empty_5_D)]{};
\node[k_ij_1, fit=(Empty_1_E) (Empty_5_E) ]{};
\node[k_ij_2, fit=(Empty_1_F) (Empty_5_F) ]{};
\end{pgfonlayer}
\end{tikzpicture}
\caption{A schematic diagram of the O2O2O2 search. As with the other diagrams, the different columns indicate different basis vectors, while different rows represent different $k_{ij}$'s. Lines indicate models produced by identical basis vectors, but different $k_{ij}$ matrices.}
\label{fig: D10_O2_L1_O2_L2_O2_L3_k_ij}
\end{figure}

\section{Conclusions} \label{sec: Conclusions}
To conclude, we decided to examine the D=10, level-1 heterotic models to deduce redundancies in the WCFFHS construction method. 
We conjectured and proved that for all models with a single basis vector, odd ordered right mover and massless left mover, the maximum number of ST SUSYs were present. 
This implies that searches of this sort in lower space-time dimensions will contain either models with the maximum number of space-time SUSYs or models without space-time supersymmetry. 
Specifically, in ten dimensions this means single basis vector searches with odd ordered right movers will only have two models: $SO(32)$ and $E_8\otimes E_8$, both with $N=1$ space-time supersymmetry. \\

For searches of two basis vector models, we showed that the basis vectors must be varied to fully map out the model spectrum. 
The GSO coefficient matrix on the other hand, does not necessarily need to be varied if the basis vectors can produce enough sets of matching boundary conditions. 
We also showed that the product of the orders across which the search is performed does not necessarily dictate the model spectrum. 
In particular, we showed that all modular invariant combinations of two order-2 basis vectors do not produce the same models as all possible order-4 basis vectors.\\

Finally, we showed that the lowest order for which all D=10, level-1 models could be produced from a single basis vector is 6. 
The lowest combination of orders which produces all of the above mentioned from pairs of basis vectors is O3O2. 
The lowest number of order-2 basis vectors needed to produce all D=10, level-1 models is three.\\

The ultimate conclusion of this study is that for simple models, it is very possible to correlate the basis vector and GSO coefficient inputs with the particle content output to a certain extent. 
These correlations make it possible to further narrow searches by analytically and statistically isolating the properties of basis vectors which give phenomenologically realistic results. 
Additional work in this area will be to explicitly map out D=8 and D=6 heterotic string models, searching for extra redundancies that occur when there are compact space-time dimensions.

\bibliography{Bibliography}

\begin{thebibliography}{66}%
\makeatletter
\providecommand \@ifxundefined [1]{%
 \@ifx{#1\undefined}
}%
\providecommand \@ifnum [1]{%
 \ifnum #1\expandafter \@firstoftwo
 \else \expandafter \@secondoftwo
 \fi
}%
\providecommand \@ifx [1]{%
 \ifx #1\expandafter \@firstoftwo
 \else \expandafter \@secondoftwo
 \fi
}%
\providecommand \natexlab [1]{#1}%
\providecommand \enquote  [1]{``#1''}%
\providecommand \bibnamefont  [1]{#1}%
\providecommand \bibfnamefont [1]{#1}%
\providecommand \citenamefont [1]{#1}%
\providecommand \href@noop [0]{\@secondoftwo}%
\providecommand \href [0]{\begingroup \@sanitize@url \@href}%
\providecommand \@href[1]{\@@startlink{#1}\@@href}%
\providecommand \@@href[1]{\endgroup#1\@@endlink}%
\providecommand \@sanitize@url [0]{\catcode `\\12\catcode `\$12\catcode
  `\&12\catcode `\#12\catcode `\^12\catcode `\_12\catcode `\%12\relax}%
\providecommand \@@startlink[1]{}%
\providecommand \@@endlink[0]{}%
\providecommand \url  [0]{\begingroup\@sanitize@url \@url }%
\providecommand \@url [1]{\endgroup\@href {#1}{\urlprefix }}%
\providecommand \urlprefix  [0]{URL }%
\providecommand \Eprint [0]{\href }%
\providecommand \doibase [0]{http://dx.doi.org/}%
\providecommand \selectlanguage [0]{\@gobble}%
\providecommand \bibinfo  [0]{\@secondoftwo}%
\providecommand \bibfield  [0]{\@secondoftwo}%
\providecommand \translation [1]{[#1]}%
\providecommand \BibitemOpen [0]{}%
\providecommand \bibitemStop [0]{}%
\providecommand \bibitemNoStop [0]{.\EOS\space}%
\providecommand \EOS [0]{\spacefactor3000\relax}%
\providecommand \BibitemShut  [1]{\csname bibitem#1\endcsname}%
\let\auto@bib@innerbib\@empty
\bibitem [{\citenamefont {Bousso}\ and\ \citenamefont
  {Polchinski}(2000)}]{Bousso:2000}%
  \BibitemOpen
  \bibfield  {author} {\bibinfo {author} {\bibfnamefont {R.}~\bibnamefont
  {Bousso}}\ and\ \bibinfo {author} {\bibfnamefont {J.}~\bibnamefont
  {Polchinski}},\ }\href@noop {} {\bibfield  {journal} {\bibinfo  {journal}
  {JHEP}\ }\textbf {\bibinfo {volume} {06}},\ \bibinfo {pages} {006} (\bibinfo
  {year} {2000})},\ \Eprint {http://arxiv.org/abs/hep-th/0004134}
  {arXiv:hep-th/0004134} \BibitemShut {NoStop}%
\bibitem [{\citenamefont {Ashok}\ and\ \citenamefont
  {Douglas}(2004)}]{Ashok:2003}%
  \BibitemOpen
  \bibfield  {author} {\bibinfo {author} {\bibfnamefont {S.}~\bibnamefont
  {Ashok}}\ and\ \bibinfo {author} {\bibfnamefont {M.~R.}\ \bibnamefont
  {Douglas}},\ }\href {\doibase 10.1088/1126-6708/2004/01/060} {\bibfield
  {journal} {\bibinfo  {journal} {JHEP}\ }\textbf {\bibinfo {volume} {01}},\
  \bibinfo {pages} {060} (\bibinfo {year} {2004})},\ \Eprint
  {http://arxiv.org/abs/hep-th/0307049} {arXiv:hep-th/0307049} \BibitemShut
  {NoStop}%
\bibitem [{\citenamefont {Donagi}\ \emph {et~al.}(2005)\citenamefont {Donagi},
  \citenamefont {He}, \citenamefont {Ovrut},\ and\ \citenamefont
  {Reinbacher}}]{Donagi:2004}%
  \BibitemOpen
  \bibfield  {author} {\bibinfo {author} {\bibfnamefont {R.}~\bibnamefont
  {Donagi}}, \bibinfo {author} {\bibfnamefont {Y.-H.}\ \bibnamefont {He}},
  \bibinfo {author} {\bibfnamefont {B.~A.}\ \bibnamefont {Ovrut}}, \ and\
  \bibinfo {author} {\bibfnamefont {R.}~\bibnamefont {Reinbacher}},\ }\href
  {\doibase 10.1088/1126-6708/2005/06/070} {\bibfield  {journal} {\bibinfo
  {journal} {JHEP}\ }\textbf {\bibinfo {volume} {06}},\ \bibinfo {pages} {070}
  (\bibinfo {year} {2005})},\ \Eprint {http://arxiv.org/abs/hep-th/0411156}
  {arXiv:hep-th/0411156} \BibitemShut {NoStop}%
\bibitem [{\citenamefont {Valandro}(2008)}]{Valandro:2008}%
  \BibitemOpen
  \bibfield  {author} {\bibinfo {author} {\bibfnamefont {R.}~\bibnamefont
  {Valandro}},\ }\href@noop {} {\  (\bibinfo {year} {2008})},\ \Eprint
  {http://arxiv.org/abs/0801.0584} {arXiv:0801.0584 [hep-th]} \BibitemShut
  {NoStop}%
\bibitem [{\citenamefont {Balasubramanian}\ \emph {et~al.}(2010)\citenamefont
  {Balasubramanian}, \citenamefont {de~Boer},\ and\ \citenamefont
  {Naqvi}}]{Balasubramanian:2008}%
  \BibitemOpen
  \bibfield  {author} {\bibinfo {author} {\bibfnamefont {V.}~\bibnamefont
  {Balasubramanian}}, \bibinfo {author} {\bibfnamefont {J.}~\bibnamefont
  {de~Boer}}, \ and\ \bibinfo {author} {\bibfnamefont {A.}~\bibnamefont
  {Naqvi}},\ }\href {\doibase 10.1016/j.physletb.2009.11.046} {\bibfield
  {journal} {\bibinfo  {journal} {Phys. Lett.}\ }\textbf {\bibinfo {volume}
  {B682}},\ \bibinfo {pages} {476} (\bibinfo {year} {2010})},\ \Eprint
  {http://arxiv.org/abs/0805.4196} {arXiv:0805.4196 [hep-th]} \BibitemShut
  {NoStop}%
\bibitem [{\citenamefont {Lebedev}\ \emph {et~al.}(2008)\citenamefont
  {Lebedev}, \citenamefont {Nilles}, \citenamefont {Ramos-Sanchez},
  \citenamefont {Ratz},\ and\ \citenamefont {Vaudrevange}}]{Lebedev:2008}%
  \BibitemOpen
  \bibfield  {author} {\bibinfo {author} {\bibfnamefont {O.}~\bibnamefont
  {Lebedev}}, \bibinfo {author} {\bibfnamefont {H.~P.}\ \bibnamefont {Nilles}},
  \bibinfo {author} {\bibfnamefont {S.}~\bibnamefont {Ramos-Sanchez}}, \bibinfo
  {author} {\bibfnamefont {M.}~\bibnamefont {Ratz}}, \ and\ \bibinfo {author}
  {\bibfnamefont {P.~K.~S.}\ \bibnamefont {Vaudrevange}},\ }\href {\doibase
  10.1016/j.physletb.2008.08.054} {\bibfield  {journal} {\bibinfo  {journal}
  {Phys. Lett.}\ }\textbf {\bibinfo {volume} {B668}},\ \bibinfo {pages} {331}
  (\bibinfo {year} {2008})},\ \Eprint {http://arxiv.org/abs/0807.4384}
  {arXiv:0807.4384 [hep-th]} \BibitemShut {NoStop}%
\bibitem [{\citenamefont {Gmeiner}\ and\ \citenamefont
  {Honecker}(2008)}]{Gmeiner:2008}%
  \BibitemOpen
  \bibfield  {author} {\bibinfo {author} {\bibfnamefont {F.}~\bibnamefont
  {Gmeiner}}\ and\ \bibinfo {author} {\bibfnamefont {G.}~\bibnamefont
  {Honecker}},\ }\href {\doibase 10.1088/1126-6708/2008/07/052} {\bibfield
  {journal} {\bibinfo  {journal} {JHEP}\ }\textbf {\bibinfo {volume} {07}},\
  \bibinfo {pages} {052} (\bibinfo {year} {2008})},\ \Eprint
  {http://arxiv.org/abs/0806.3039} {arXiv:0806.3039 [hep-th]} \BibitemShut
  {NoStop}%
\bibitem [{\citenamefont {Dienes}\ and\ \citenamefont
  {Lennek}(2009)}]{Dienes:2008}%
  \BibitemOpen
  \bibfield  {author} {\bibinfo {author} {\bibfnamefont {K.~R.}\ \bibnamefont
  {Dienes}}\ and\ \bibinfo {author} {\bibfnamefont {M.}~\bibnamefont
  {Lennek}},\ }\href {\doibase 10.1103/PhysRevD.80.106003} {\bibfield
  {journal} {\bibinfo  {journal} {Phys. Rev.}\ }\textbf {\bibinfo {volume}
  {D80}},\ \bibinfo {pages} {106003} (\bibinfo {year} {2009})},\ \Eprint
  {http://arxiv.org/abs/0809.0036} {arXiv:0809.0036 [hep-th]} \BibitemShut
  {NoStop}%
\bibitem [{\citenamefont {Gabella}\ \emph {et~al.}(2008)\citenamefont
  {Gabella}, \citenamefont {He},\ and\ \citenamefont {Lukas}}]{Gabella:2008}%
  \BibitemOpen
  \bibfield  {author} {\bibinfo {author} {\bibfnamefont {M.}~\bibnamefont
  {Gabella}}, \bibinfo {author} {\bibfnamefont {Y.-H.}\ \bibnamefont {He}}, \
  and\ \bibinfo {author} {\bibfnamefont {A.}~\bibnamefont {Lukas}},\ }\href
  {\doibase 10.1088/1126-6708/2008/12/027} {\bibfield  {journal} {\bibinfo
  {journal} {JHEP}\ }\textbf {\bibinfo {volume} {12}},\ \bibinfo {pages} {027}
  (\bibinfo {year} {2008})},\ \Eprint {http://arxiv.org/abs/0808.2142}
  {arXiv:0808.2142 [hep-th]} \BibitemShut {NoStop}%
\bibitem [{\citenamefont {Donagi}\ and\ \citenamefont
  {Wendland}(2009)}]{Donagi:2008}%
  \BibitemOpen
  \bibfield  {author} {\bibinfo {author} {\bibfnamefont {R.}~\bibnamefont
  {Donagi}}\ and\ \bibinfo {author} {\bibfnamefont {K.}~\bibnamefont
  {Wendland}},\ }\href {\doibase 10.1016/j.geomphys.2009.04.004} {\bibfield
  {journal} {\bibinfo  {journal} {J. Geom. Phys.}\ }\textbf {\bibinfo {volume}
  {59}},\ \bibinfo {pages} {942} (\bibinfo {year} {2009})},\ \Eprint
  {http://arxiv.org/abs/0809.0330} {arXiv:0809.0330 [hep-th]} \BibitemShut
  {NoStop}%
\bibitem [{\citenamefont {Antoniadis}\ \emph {et~al.}(1987)\citenamefont
  {Antoniadis}, \citenamefont {Bachas},\ and\ \citenamefont
  {Kounnas}}]{Antoniadis:1986}%
  \BibitemOpen
  \bibfield  {author} {\bibinfo {author} {\bibfnamefont {I.}~\bibnamefont
  {Antoniadis}}, \bibinfo {author} {\bibfnamefont {C.~P.}\ \bibnamefont
  {Bachas}}, \ and\ \bibinfo {author} {\bibfnamefont {C.}~\bibnamefont
  {Kounnas}},\ }\href {\doibase 10.1016/0550-3213(87)90372-5} {\bibfield
  {journal} {\bibinfo  {journal} {Nucl. Phys.}\ }\textbf {\bibinfo {volume}
  {B289}},\ \bibinfo {pages} {87} (\bibinfo {year} {1987})}\BibitemShut
  {NoStop}%
\bibitem [{\citenamefont {Antoniadis}\ and\ \citenamefont
  {Bachas}(1988)}]{Antoniadis:1987}%
  \BibitemOpen
  \bibfield  {author} {\bibinfo {author} {\bibfnamefont {I.}~\bibnamefont
  {Antoniadis}}\ and\ \bibinfo {author} {\bibfnamefont {C.}~\bibnamefont
  {Bachas}},\ }\href {\doibase 10.1016/0550-3213(88)90355-0} {\bibfield
  {journal} {\bibinfo  {journal} {Nucl. Phys.}\ }\textbf {\bibinfo {volume}
  {B298}},\ \bibinfo {pages} {586} (\bibinfo {year} {1988})}\BibitemShut
  {NoStop}%
\bibitem [{\citenamefont {Kawai}\ \emph {et~al.}(1987)\citenamefont {Kawai},
  \citenamefont {Lewellen},\ and\ \citenamefont {Tye}}]{Kawai:1986_2}%
  \BibitemOpen
  \bibfield  {author} {\bibinfo {author} {\bibfnamefont {H.}~\bibnamefont
  {Kawai}}, \bibinfo {author} {\bibfnamefont {D.~C.}\ \bibnamefont {Lewellen}},
  \ and\ \bibinfo {author} {\bibfnamefont {S.~H.~H.}\ \bibnamefont {Tye}},\
  }\href {\doibase 10.1016/0550-3213(87)90208-2} {\bibfield  {journal}
  {\bibinfo  {journal} {Nucl. Phys.}\ }\textbf {\bibinfo {volume} {B288}},\
  \bibinfo {pages} {1} (\bibinfo {year} {1987})}\BibitemShut {NoStop}%
\bibitem [{\citenamefont {Cleaver}\ \emph
  {et~al.}(2001{\natexlab{a}})\citenamefont {Cleaver}, \citenamefont {Faraggi},
  \citenamefont {Nanopoulos},\ and\ \citenamefont {Walker}}]{Cleaver:1999}%
  \BibitemOpen
  \bibfield  {author} {\bibinfo {author} {\bibfnamefont {G.~B.}\ \bibnamefont
  {Cleaver}}, \bibinfo {author} {\bibfnamefont {A.~E.}\ \bibnamefont
  {Faraggi}}, \bibinfo {author} {\bibfnamefont {D.~V.}\ \bibnamefont
  {Nanopoulos}}, \ and\ \bibinfo {author} {\bibfnamefont {J.~W.}\ \bibnamefont
  {Walker}},\ }\href {\doibase 10.1016/S0550-3213(00)00543-5} {\bibfield
  {journal} {\bibinfo  {journal} {Nucl. Phys.}\ }\textbf {\bibinfo {volume}
  {B593}},\ \bibinfo {pages} {471} (\bibinfo {year} {2001}{\natexlab{a}})},\
  \Eprint {http://arxiv.org/abs/hep-ph/9910230} {arXiv:hep-ph/9910230}
  \BibitemShut {NoStop}%
\bibitem [{\citenamefont {Lopez}\ \emph {et~al.}(1993)\citenamefont {Lopez},
  \citenamefont {Nanopoulos},\ and\ \citenamefont {Yuan}}]{Lopez:1992}%
  \BibitemOpen
  \bibfield  {author} {\bibinfo {author} {\bibfnamefont {J.~L.}\ \bibnamefont
  {Lopez}}, \bibinfo {author} {\bibfnamefont {D.~V.}\ \bibnamefont
  {Nanopoulos}}, \ and\ \bibinfo {author} {\bibfnamefont {K.-j.}\ \bibnamefont
  {Yuan}},\ }\href {\doibase 10.1016/0550-3213(93)90513-O} {\bibfield
  {journal} {\bibinfo  {journal} {Nucl. Phys.}\ }\textbf {\bibinfo {volume}
  {B399}},\ \bibinfo {pages} {654} (\bibinfo {year} {1993})},\ \Eprint
  {http://arxiv.org/abs/hep-th/9203025} {arXiv:hep-th/9203025} \BibitemShut
  {NoStop}%
\bibitem [{\citenamefont {Faraggi}\ \emph {et~al.}(1990)\citenamefont
  {Faraggi}, \citenamefont {Nanopoulos},\ and\ \citenamefont
  {Yuan}}]{Faraggi:1989}%
  \BibitemOpen
  \bibfield  {author} {\bibinfo {author} {\bibfnamefont {A.~E.}\ \bibnamefont
  {Faraggi}}, \bibinfo {author} {\bibfnamefont {D.~V.}\ \bibnamefont
  {Nanopoulos}}, \ and\ \bibinfo {author} {\bibfnamefont {K.-j.}\ \bibnamefont
  {Yuan}},\ }\href {\doibase 10.1016/0550-3213(90)90498-3} {\bibfield
  {journal} {\bibinfo  {journal} {Nucl. Phys.}\ }\textbf {\bibinfo {volume}
  {B335}},\ \bibinfo {pages} {347} (\bibinfo {year} {1990})}\BibitemShut
  {NoStop}%
\bibitem [{\citenamefont {Faraggi}(1992{\natexlab{a}})}]{Faraggi:1992}%
  \BibitemOpen
  \bibfield  {author} {\bibinfo {author} {\bibfnamefont {A.~E.}\ \bibnamefont
  {Faraggi}},\ }\href {\doibase 10.1016/0550-3213(92)90160-D} {\bibfield
  {journal} {\bibinfo  {journal} {Nucl. Phys.}\ }\textbf {\bibinfo {volume}
  {B387}},\ \bibinfo {pages} {239} (\bibinfo {year} {1992}{\natexlab{a}})},\
  \Eprint {http://arxiv.org/abs/hep-th/9208024} {arXiv:hep-th/9208024}
  \BibitemShut {NoStop}%
\bibitem [{\citenamefont {Antoniadis}\ \emph {et~al.}(1990)\citenamefont
  {Antoniadis}, \citenamefont {Leontaris},\ and\ \citenamefont
  {Rizos}}]{Antoniadis:1990}%
  \BibitemOpen
  \bibfield  {author} {\bibinfo {author} {\bibfnamefont {I.}~\bibnamefont
  {Antoniadis}}, \bibinfo {author} {\bibfnamefont {G.~K.}\ \bibnamefont
  {Leontaris}}, \ and\ \bibinfo {author} {\bibfnamefont {J.}~\bibnamefont
  {Rizos}},\ }\href {\doibase 10.1016/0370-2693(90)90127-R} {\bibfield
  {journal} {\bibinfo  {journal} {Phys. Lett.}\ }\textbf {\bibinfo {volume}
  {B245}},\ \bibinfo {pages} {161} (\bibinfo {year} {1990})}\BibitemShut
  {NoStop}%
\bibitem [{\citenamefont {Leontaris}\ and\ \citenamefont
  {Rizos}(1999)}]{Leontaris:1999}%
  \BibitemOpen
  \bibfield  {author} {\bibinfo {author} {\bibfnamefont {G.~K.}\ \bibnamefont
  {Leontaris}}\ and\ \bibinfo {author} {\bibfnamefont {J.}~\bibnamefont
  {Rizos}},\ }\href {\doibase 10.1016/S0550-3213(99)00303-X} {\bibfield
  {journal} {\bibinfo  {journal} {Nucl. Phys.}\ }\textbf {\bibinfo {volume}
  {B554}},\ \bibinfo {pages} {3} (\bibinfo {year} {1999})},\ \Eprint
  {http://arxiv.org/abs/hep-th/9901098} {arXiv:hep-th/9901098} \BibitemShut
  {NoStop}%
\bibitem [{\citenamefont {Faraggi}(1992{\natexlab{b}})}]{Faraggi:1991}%
  \BibitemOpen
  \bibfield  {author} {\bibinfo {author} {\bibfnamefont {A.~E.}\ \bibnamefont
  {Faraggi}},\ }\href {\doibase 10.1016/0370-2693(92)90723-H} {\bibfield
  {journal} {\bibinfo  {journal} {Phys. Lett.}\ }\textbf {\bibinfo {volume}
  {B278}},\ \bibinfo {pages} {131} (\bibinfo {year}
  {1992}{\natexlab{b}})}\BibitemShut {NoStop}%
\bibitem [{\citenamefont {Faraggi}(1993{\natexlab{a}})}]{Faraggi:1992_2}%
  \BibitemOpen
  \bibfield  {author} {\bibinfo {author} {\bibfnamefont {A.~E.}\ \bibnamefont
  {Faraggi}},\ }\href {\doibase 10.1016/0550-3213(93)90030-S} {\bibfield
  {journal} {\bibinfo  {journal} {Nucl. Phys.}\ }\textbf {\bibinfo {volume}
  {B403}},\ \bibinfo {pages} {101} (\bibinfo {year} {1993}{\natexlab{a}})},\
  \Eprint {http://arxiv.org/abs/hep-th/9208023} {arXiv:hep-th/9208023}
  \BibitemShut {NoStop}%
\bibitem [{\citenamefont {Faraggi}(1993{\natexlab{b}})}]{Faraggi:1992_3}%
  \BibitemOpen
  \bibfield  {author} {\bibinfo {author} {\bibfnamefont {A.~E.}\ \bibnamefont
  {Faraggi}},\ }\href {\doibase 10.1016/0550-3213(93)90273-R} {\bibfield
  {journal} {\bibinfo  {journal} {Nucl. Phys.}\ }\textbf {\bibinfo {volume}
  {B407}},\ \bibinfo {pages} {57} (\bibinfo {year} {1993}{\natexlab{b}})},\
  \Eprint {http://arxiv.org/abs/hep-ph/9210256} {arXiv:hep-ph/9210256}
  \BibitemShut {NoStop}%
\bibitem [{\citenamefont {Faraggi}(1992{\natexlab{c}})}]{Faraggi:1991_2}%
  \BibitemOpen
  \bibfield  {author} {\bibinfo {author} {\bibfnamefont {A.~E.}\ \bibnamefont
  {Faraggi}},\ }\href {\doibase 10.1016/0370-2693(92)90302-K} {\bibfield
  {journal} {\bibinfo  {journal} {Phys. Lett.}\ }\textbf {\bibinfo {volume}
  {B274}},\ \bibinfo {pages} {47} (\bibinfo {year}
  {1992}{\natexlab{c}})}\BibitemShut {NoStop}%
\bibitem [{\citenamefont {Faraggi}(1993{\natexlab{c}})}]{Faraggi:1991_3}%
  \BibitemOpen
  \bibfield  {author} {\bibinfo {author} {\bibfnamefont {A.~E.}\ \bibnamefont
  {Faraggi}},\ }\href {\doibase 10.1103/PhysRevD.47.5021} {\bibfield  {journal}
  {\bibinfo  {journal} {Phys. Rev.}\ }\textbf {\bibinfo {volume} {D47}},\
  \bibinfo {pages} {5021} (\bibinfo {year} {1993}{\natexlab{c}})}\BibitemShut
  {NoStop}%
\bibitem [{\citenamefont {Faraggi}(1996)}]{Faraggi:1995}%
  \BibitemOpen
  \bibfield  {author} {\bibinfo {author} {\bibfnamefont {A.~E.}\ \bibnamefont
  {Faraggi}},\ }\href {\doibase 10.1016/0370-2693(96)00310-3} {\bibfield
  {journal} {\bibinfo  {journal} {Phys. Lett.}\ }\textbf {\bibinfo {volume}
  {B377}},\ \bibinfo {pages} {43} (\bibinfo {year} {1996})},\ \Eprint
  {http://arxiv.org/abs/hep-ph/9506388} {arXiv:hep-ph/9506388} \BibitemShut
  {NoStop}%
\bibitem [{\citenamefont {Faraggi}(1997)}]{Faraggi:1996}%
  \BibitemOpen
  \bibfield  {author} {\bibinfo {author} {\bibfnamefont {A.~E.}\ \bibnamefont
  {Faraggi}},\ }\href {\doibase 10.1016/S0550-3213(96)00682-7} {\bibfield
  {journal} {\bibinfo  {journal} {Nucl. Phys.}\ }\textbf {\bibinfo {volume}
  {B487}},\ \bibinfo {pages} {55} (\bibinfo {year} {1997})},\ \Eprint
  {http://arxiv.org/abs/hep-ph/9601332} {arXiv:hep-ph/9601332} \BibitemShut
  {NoStop}%
\bibitem [{\citenamefont {Cleaver}(1998{\natexlab{a}})}]{Cleaver:1997}%
  \BibitemOpen
  \bibfield  {author} {\bibinfo {author} {\bibfnamefont {G.~B.}\ \bibnamefont
  {Cleaver}},\ }\href {\doibase 10.1016/S0920-5632(97)00653-1} {\bibfield
  {journal} {\bibinfo  {journal} {Nucl. Phys. Proc. Suppl.}\ }\textbf {\bibinfo
  {volume} {62}},\ \bibinfo {pages} {161} (\bibinfo {year}
  {1998}{\natexlab{a}})},\ \Eprint {http://arxiv.org/abs/hep-th/9708023}
  {arXiv:hep-th/9708023} \BibitemShut {NoStop}%
\bibitem [{\citenamefont {Cleaver}\ and\ \citenamefont
  {Faraggi}(1999)}]{Cleaver:1997_2}%
  \BibitemOpen
  \bibfield  {author} {\bibinfo {author} {\bibfnamefont {G.~B.}\ \bibnamefont
  {Cleaver}}\ and\ \bibinfo {author} {\bibfnamefont {A.~E.}\ \bibnamefont
  {Faraggi}},\ }\href {\doibase 10.1142/S0217751X99001172} {\bibfield
  {journal} {\bibinfo  {journal} {Int. J. Mod. Phys.}\ }\textbf {\bibinfo
  {volume} {A14}},\ \bibinfo {pages} {2335} (\bibinfo {year} {1999})},\ \Eprint
  {http://arxiv.org/abs/hep-ph/9711339} {arXiv:hep-ph/9711339} \BibitemShut
  {NoStop}%
\bibitem [{\citenamefont {Cleaver}\ \emph {et~al.}(1998)\citenamefont
  {Cleaver}, \citenamefont {Cvetic}, \citenamefont {Espinosa}, \citenamefont
  {Everett},\ and\ \citenamefont {Langacker}}]{Cleaver:1997_3}%
  \BibitemOpen
  \bibfield  {author} {\bibinfo {author} {\bibfnamefont {G.}~\bibnamefont
  {Cleaver}}, \bibinfo {author} {\bibfnamefont {M.}~\bibnamefont {Cvetic}},
  \bibinfo {author} {\bibfnamefont {J.~R.}\ \bibnamefont {Espinosa}}, \bibinfo
  {author} {\bibfnamefont {L.~L.}\ \bibnamefont {Everett}}, \ and\ \bibinfo
  {author} {\bibfnamefont {P.}~\bibnamefont {Langacker}},\ }\href {\doibase
  10.1016/S0550-3213(98)00277-6} {\bibfield  {journal} {\bibinfo  {journal}
  {Nucl. Phys.}\ }\textbf {\bibinfo {volume} {B525}},\ \bibinfo {pages} {3}
  (\bibinfo {year} {1998})},\ \Eprint {http://arxiv.org/abs/hep-th/9711178}
  {arXiv:hep-th/9711178} \BibitemShut {NoStop}%
\bibitem [{\citenamefont {Cleaver}\ \emph
  {et~al.}(1999{\natexlab{a}})\citenamefont {Cleaver}, \citenamefont {Cvetic},
  \citenamefont {Espinosa}, \citenamefont {Everett},\ and\ \citenamefont
  {Langacker}}]{Cleaver:1998}%
  \BibitemOpen
  \bibfield  {author} {\bibinfo {author} {\bibfnamefont {G.}~\bibnamefont
  {Cleaver}}, \bibinfo {author} {\bibfnamefont {M.}~\bibnamefont {Cvetic}},
  \bibinfo {author} {\bibfnamefont {J.~R.}\ \bibnamefont {Espinosa}}, \bibinfo
  {author} {\bibfnamefont {L.~L.}\ \bibnamefont {Everett}}, \ and\ \bibinfo
  {author} {\bibfnamefont {P.}~\bibnamefont {Langacker}},\ }\href {\doibase
  10.1016/S0550-3213(98)00863-3} {\bibfield  {journal} {\bibinfo  {journal}
  {Nucl. Phys.}\ }\textbf {\bibinfo {volume} {B545}},\ \bibinfo {pages} {47}
  (\bibinfo {year} {1999}{\natexlab{a}})},\ \Eprint
  {http://arxiv.org/abs/hep-th/9805133} {arXiv:hep-th/9805133} \BibitemShut
  {NoStop}%
\bibitem [{\citenamefont {Cleaver}\ \emph
  {et~al.}(1999{\natexlab{b}})\citenamefont {Cleaver} \emph
  {et~al.}}]{Cleaver:1998_2}%
  \BibitemOpen
  \bibfield  {author} {\bibinfo {author} {\bibfnamefont {G.}~\bibnamefont
  {Cleaver}} \emph {et~al.},\ }\href {\doibase 10.1103/PhysRevD.59.055005}
  {\bibfield  {journal} {\bibinfo  {journal} {Phys. Rev.}\ }\textbf {\bibinfo
  {volume} {D59}},\ \bibinfo {pages} {055005} (\bibinfo {year}
  {1999}{\natexlab{b}})},\ \Eprint {http://arxiv.org/abs/hep-ph/9807479}
  {arXiv:hep-ph/9807479} \BibitemShut {NoStop}%
\bibitem [{\citenamefont {Cleaver}\ \emph
  {et~al.}(1999{\natexlab{c}})\citenamefont {Cleaver} \emph
  {et~al.}}]{Cleaver:1998_3}%
  \BibitemOpen
  \bibfield  {author} {\bibinfo {author} {\bibfnamefont {G.}~\bibnamefont
  {Cleaver}} \emph {et~al.},\ }\href {\doibase 10.1103/PhysRevD.59.115003}
  {\bibfield  {journal} {\bibinfo  {journal} {Phys. Rev.}\ }\textbf {\bibinfo
  {volume} {D59}},\ \bibinfo {pages} {115003} (\bibinfo {year}
  {1999}{\natexlab{c}})},\ \Eprint {http://arxiv.org/abs/hep-ph/9811355}
  {arXiv:hep-ph/9811355} \BibitemShut {NoStop}%
\bibitem [{\citenamefont {Cleaver}(1998{\natexlab{b}})}]{Cleaver:1998_4}%
  \BibitemOpen
  \bibfield  {author} {\bibinfo {author} {\bibfnamefont {G.~B.}\ \bibnamefont
  {Cleaver}},\ }\href@noop {} {\  (\bibinfo {year} {1998}{\natexlab{b}})},\
  \Eprint {http://arxiv.org/abs/hep-ph/9812262} {arXiv:hep-ph/9812262}
  \BibitemShut {NoStop}%
\bibitem [{\citenamefont {Cleaver}\ \emph
  {et~al.}(1999{\natexlab{d}})\citenamefont {Cleaver}, \citenamefont
  {Faraggi},\ and\ \citenamefont {Nanopoulos}}]{Cleaver:1998_5}%
  \BibitemOpen
  \bibfield  {author} {\bibinfo {author} {\bibfnamefont {G.~B.}\ \bibnamefont
  {Cleaver}}, \bibinfo {author} {\bibfnamefont {A.~E.}\ \bibnamefont
  {Faraggi}}, \ and\ \bibinfo {author} {\bibfnamefont {D.~V.}\ \bibnamefont
  {Nanopoulos}},\ }\href {\doibase 10.1016/S0370-2693(99)00413-X} {\bibfield
  {journal} {\bibinfo  {journal} {Phys. Lett.}\ }\textbf {\bibinfo {volume}
  {B455}},\ \bibinfo {pages} {135} (\bibinfo {year} {1999}{\natexlab{d}})},\
  \Eprint {http://arxiv.org/abs/hep-ph/9811427} {arXiv:hep-ph/9811427}
  \BibitemShut {NoStop}%
\bibitem [{\citenamefont {Cleaver}\ \emph
  {et~al.}(2001{\natexlab{b}})\citenamefont {Cleaver}, \citenamefont
  {Faraggi},\ and\ \citenamefont {Nanopoulos}}]{Cleaver:1999_2}%
  \BibitemOpen
  \bibfield  {author} {\bibinfo {author} {\bibfnamefont {G.~B.}\ \bibnamefont
  {Cleaver}}, \bibinfo {author} {\bibfnamefont {A.~E.}\ \bibnamefont
  {Faraggi}}, \ and\ \bibinfo {author} {\bibfnamefont {D.~V.}\ \bibnamefont
  {Nanopoulos}},\ }\href {\doibase 10.1142/S0217751X01001057} {\bibfield
  {journal} {\bibinfo  {journal} {Int. J. Mod. Phys.}\ }\textbf {\bibinfo
  {volume} {A16}},\ \bibinfo {pages} {425} (\bibinfo {year}
  {2001}{\natexlab{b}})},\ \Eprint {http://arxiv.org/abs/hep-ph/9904301}
  {arXiv:hep-ph/9904301} \BibitemShut {NoStop}%
\bibitem [{\citenamefont {Cleaver}\ \emph
  {et~al.}(2001{\natexlab{c}})\citenamefont {Cleaver}, \citenamefont {Faraggi},
  \citenamefont {Nanopoulos},\ and\ \citenamefont {Walker}}]{Cleaver:1999_3}%
  \BibitemOpen
  \bibfield  {author} {\bibinfo {author} {\bibfnamefont {G.~B.}\ \bibnamefont
  {Cleaver}}, \bibinfo {author} {\bibfnamefont {A.~E.}\ \bibnamefont
  {Faraggi}}, \bibinfo {author} {\bibfnamefont {D.~V.}\ \bibnamefont
  {Nanopoulos}}, \ and\ \bibinfo {author} {\bibfnamefont {J.~W.}\ \bibnamefont
  {Walker}},\ }\href {\doibase 10.1016/S0550-3213(00)00543-5} {\bibfield
  {journal} {\bibinfo  {journal} {Nucl. Phys.}\ }\textbf {\bibinfo {volume}
  {B593}},\ \bibinfo {pages} {471} (\bibinfo {year} {2001}{\natexlab{c}})},\
  \Eprint {http://arxiv.org/abs/hep-ph/9910230} {arXiv:hep-ph/9910230}
  \BibitemShut {NoStop}%
\bibitem [{\citenamefont {Cleaver}(1999)}]{Cleaver:1999_4}%
  \BibitemOpen
  \bibfield  {author} {\bibinfo {author} {\bibfnamefont {G.~B.}\ \bibnamefont
  {Cleaver}},\ }\href@noop {} {\  (\bibinfo {year} {1999})},\ \Eprint
  {http://arxiv.org/abs/hep-ph/9901203} {arXiv:hep-ph/9901203} \BibitemShut
  {NoStop}%
\bibitem [{\citenamefont {Cleaver}\ \emph {et~al.}(2000)\citenamefont
  {Cleaver}, \citenamefont {Faraggi}, \citenamefont {Nanopoulos},\ and\
  \citenamefont {Walker}}]{Cleaver:2000}%
  \BibitemOpen
  \bibfield  {author} {\bibinfo {author} {\bibfnamefont {G.~B.}\ \bibnamefont
  {Cleaver}}, \bibinfo {author} {\bibfnamefont {A.~E.}\ \bibnamefont
  {Faraggi}}, \bibinfo {author} {\bibfnamefont {D.~V.}\ \bibnamefont
  {Nanopoulos}}, \ and\ \bibinfo {author} {\bibfnamefont {J.~W.}\ \bibnamefont
  {Walker}},\ }\href {\doibase 10.1142/S0217732300001444} {\bibfield  {journal}
  {\bibinfo  {journal} {Mod. Phys. Lett.}\ }\textbf {\bibinfo {volume} {A15}},\
  \bibinfo {pages} {1191} (\bibinfo {year} {2000})},\ \Eprint
  {http://arxiv.org/abs/hep-ph/0002060} {arXiv:hep-ph/0002060} \BibitemShut
  {NoStop}%
\bibitem [{\citenamefont {Cleaver}\ \emph
  {et~al.}(2001{\natexlab{d}})\citenamefont {Cleaver}, \citenamefont
  {Faraggi},\ and\ \citenamefont {Savage}}]{Cleaver:2000_2}%
  \BibitemOpen
  \bibfield  {author} {\bibinfo {author} {\bibfnamefont {G.~B.}\ \bibnamefont
  {Cleaver}}, \bibinfo {author} {\bibfnamefont {A.~E.}\ \bibnamefont
  {Faraggi}}, \ and\ \bibinfo {author} {\bibfnamefont {C.}~\bibnamefont
  {Savage}},\ }\href {\doibase 10.1103/PhysRevD.63.066001} {\bibfield
  {journal} {\bibinfo  {journal} {Phys. Rev.}\ }\textbf {\bibinfo {volume}
  {D63}},\ \bibinfo {pages} {066001} (\bibinfo {year} {2001}{\natexlab{d}})},\
  \Eprint {http://arxiv.org/abs/hep-ph/0006331} {arXiv:hep-ph/0006331}
  \BibitemShut {NoStop}%
\bibitem [{\citenamefont {Cleaver}\ \emph
  {et~al.}(2002{\natexlab{a}})\citenamefont {Cleaver}, \citenamefont {Faraggi},
  \citenamefont {Nanopoulos},\ and\ \citenamefont {Walker}}]{Cleaver:2001}%
  \BibitemOpen
  \bibfield  {author} {\bibinfo {author} {\bibfnamefont {G.~B.}\ \bibnamefont
  {Cleaver}}, \bibinfo {author} {\bibfnamefont {A.~E.}\ \bibnamefont
  {Faraggi}}, \bibinfo {author} {\bibfnamefont {D.~V.}\ \bibnamefont
  {Nanopoulos}}, \ and\ \bibinfo {author} {\bibfnamefont {J.~W.}\ \bibnamefont
  {Walker}},\ }\href {\doibase 10.1016/S0550-3213(01)00558-2} {\bibfield
  {journal} {\bibinfo  {journal} {Nucl. Phys.}\ }\textbf {\bibinfo {volume}
  {B620}},\ \bibinfo {pages} {259} (\bibinfo {year} {2002}{\natexlab{a}})},\
  \Eprint {http://arxiv.org/abs/hep-ph/0104091} {arXiv:hep-ph/0104091}
  \BibitemShut {NoStop}%
\bibitem [{\citenamefont {Cleaver}\ \emph
  {et~al.}(2002{\natexlab{b}})\citenamefont {Cleaver}, \citenamefont
  {Clements},\ and\ \citenamefont {Faraggi}}]{Cleaver:2001_2}%
  \BibitemOpen
  \bibfield  {author} {\bibinfo {author} {\bibfnamefont {G.~B.}\ \bibnamefont
  {Cleaver}}, \bibinfo {author} {\bibfnamefont {D.~J.}\ \bibnamefont
  {Clements}}, \ and\ \bibinfo {author} {\bibfnamefont {A.~E.}\ \bibnamefont
  {Faraggi}},\ }\href {\doibase 10.1103/PhysRevD.65.106003} {\bibfield
  {journal} {\bibinfo  {journal} {Phys. Rev.}\ }\textbf {\bibinfo {volume}
  {D65}},\ \bibinfo {pages} {106003} (\bibinfo {year} {2002}{\natexlab{b}})},\
  \Eprint {http://arxiv.org/abs/hep-ph/0106060} {arXiv:hep-ph/0106060}
  \BibitemShut {NoStop}%
\bibitem [{\citenamefont {Cleaver}\ \emph
  {et~al.}(2003{\natexlab{a}})\citenamefont {Cleaver}, \citenamefont
  {Faraggi},\ and\ \citenamefont {Nooij}}]{Cleaver:2002}%
  \BibitemOpen
  \bibfield  {author} {\bibinfo {author} {\bibfnamefont {G.~B.}\ \bibnamefont
  {Cleaver}}, \bibinfo {author} {\bibfnamefont {A.~E.}\ \bibnamefont
  {Faraggi}}, \ and\ \bibinfo {author} {\bibfnamefont {S.}~\bibnamefont
  {Nooij}},\ }\href {\doibase 10.1016/j.nuclphysb.2003.09.012} {\bibfield
  {journal} {\bibinfo  {journal} {Nucl. Phys.}\ }\textbf {\bibinfo {volume}
  {B672}},\ \bibinfo {pages} {64} (\bibinfo {year} {2003}{\natexlab{a}})},\
  \Eprint {http://arxiv.org/abs/hep-ph/0301037} {arXiv:hep-ph/0301037}
  \BibitemShut {NoStop}%
\bibitem [{\citenamefont {Cleaver}(2002)}]{Cleaver:2002_2}%
  \BibitemOpen
  \bibfield  {author} {\bibinfo {author} {\bibfnamefont {G.~B.}\ \bibnamefont
  {Cleaver}},\ }\href@noop {} {\  (\bibinfo {year} {2002})},\ \Eprint
  {http://arxiv.org/abs/hep-ph/0210093} {arXiv:hep-ph/0210093} \BibitemShut
  {NoStop}%
\bibitem [{\citenamefont {Cleaver}\ \emph
  {et~al.}(2003{\natexlab{b}})\citenamefont {Cleaver} \emph
  {et~al.}}]{Cleaver:2002_3}%
  \BibitemOpen
  \bibfield  {author} {\bibinfo {author} {\bibfnamefont {G.}~\bibnamefont
  {Cleaver}} \emph {et~al.},\ }\href {\doibase 10.1103/PhysRevD.67.026009}
  {\bibfield  {journal} {\bibinfo  {journal} {Phys. Rev.}\ }\textbf {\bibinfo
  {volume} {D67}},\ \bibinfo {pages} {026009} (\bibinfo {year}
  {2003}{\natexlab{b}})},\ \Eprint {http://arxiv.org/abs/hep-ph/0209050}
  {arXiv:hep-ph/0209050} \BibitemShut {NoStop}%
\bibitem [{\citenamefont {Perkins}\ \emph {et~al.}(2003)\citenamefont {Perkins}
  \emph {et~al.}}]{Perkins:2003}%
  \BibitemOpen
  \bibfield  {author} {\bibinfo {author} {\bibfnamefont {J.}~\bibnamefont
  {Perkins}} \emph {et~al.},\ }\href@noop {} {\  (\bibinfo {year} {2003})},\
  \Eprint {http://arxiv.org/abs/hep-ph/0310155} {arXiv:hep-ph/0310155}
  \BibitemShut {NoStop}%
\bibitem [{\citenamefont {Perkins}\ \emph {et~al.}(2007)\citenamefont {Perkins}
  \emph {et~al.}}]{Perkins:2005}%
  \BibitemOpen
  \bibfield  {author} {\bibinfo {author} {\bibfnamefont {J.}~\bibnamefont
  {Perkins}} \emph {et~al.},\ }\href {\doibase 10.1103/PhysRevD.75.026007}
  {\bibfield  {journal} {\bibinfo  {journal} {Phys. Rev.}\ }\textbf {\bibinfo
  {volume} {D75}},\ \bibinfo {pages} {026007} (\bibinfo {year} {2007})},\
  \Eprint {http://arxiv.org/abs/hep-ph/0510141} {arXiv:hep-ph/0510141}
  \BibitemShut {NoStop}%
\bibitem [{\citenamefont {Cleaver}\ \emph {et~al.}(2008)\citenamefont
  {Cleaver}, \citenamefont {Faraggi}, \citenamefont {Manno},\ and\
  \citenamefont {Timirgaziu}}]{Cleaver:2008}%
  \BibitemOpen
  \bibfield  {author} {\bibinfo {author} {\bibfnamefont {G.~B.}\ \bibnamefont
  {Cleaver}}, \bibinfo {author} {\bibfnamefont {A.~E.}\ \bibnamefont
  {Faraggi}}, \bibinfo {author} {\bibfnamefont {E.}~\bibnamefont {Manno}}, \
  and\ \bibinfo {author} {\bibfnamefont {C.}~\bibnamefont {Timirgaziu}},\
  }\href {\doibase 10.1103/PhysRevD.78.046009} {\bibfield  {journal} {\bibinfo
  {journal} {Phys. Rev.}\ }\textbf {\bibinfo {volume} {D78}},\ \bibinfo {pages}
  {046009} (\bibinfo {year} {2008})},\ \Eprint {http://arxiv.org/abs/0802.0470}
  {arXiv:0802.0470 [hep-th]} \BibitemShut {NoStop}%
\bibitem [{\citenamefont {Greenwald}\ \emph {et~al.}(2009)\citenamefont
  {Greenwald}, \citenamefont {Pechan}, \citenamefont {Renner}, \citenamefont
  {Ali},\ and\ \citenamefont {Cleaver}}]{Greenwald:2009}%
  \BibitemOpen
  \bibfield  {author} {\bibinfo {author} {\bibfnamefont {D.}~\bibnamefont
  {Greenwald}, \bibfnamefont {Jared~Moore}}, \bibinfo {author} {\bibfnamefont
  {K.}~\bibnamefont {Pechan}}, \bibinfo {author} {\bibfnamefont
  {T.}~\bibnamefont {Renner}}, \bibinfo {author} {\bibfnamefont
  {T.}~\bibnamefont {Ali}}, \ and\ \bibinfo {author} {\bibfnamefont
  {G.}~\bibnamefont {Cleaver}},\ }\href@noop {} {\  (\bibinfo {year} {2009})},\
  \Eprint {http://arxiv.org/abs/0912.5207} {arXiv:0912.5207 [hep-ph]}
  \BibitemShut {NoStop}%
\bibitem [{\citenamefont {Cleaver}\ \emph {et~al.}(2011)\citenamefont {Cleaver}
  \emph {et~al.}}]{Cleaver:2011}%
  \BibitemOpen
  \bibfield  {author} {\bibinfo {author} {\bibfnamefont {G.}~\bibnamefont
  {Cleaver}} \emph {et~al.},\ }\href@noop {} {\  (\bibinfo {year} {2011})},\
  \Eprint {http://arxiv.org/abs/1105.0447} {arXiv:1105.0447 [hep-ph]}
  \BibitemShut {NoStop}%
\bibitem [{\citenamefont {Dienes}(2006)}]{Dienes:2006}%
  \BibitemOpen
  \bibfield  {author} {\bibinfo {author} {\bibfnamefont {K.~R.}\ \bibnamefont
  {Dienes}},\ }\href {\doibase 10.1103/PhysRevD.73.106010} {\bibfield
  {journal} {\bibinfo  {journal} {Phys. Rev.}\ }\textbf {\bibinfo {volume}
  {D73}},\ \bibinfo {pages} {106010} (\bibinfo {year} {2006})},\ \Eprint
  {http://arxiv.org/abs/hep-th/0602286} {arXiv:hep-th/0602286} \BibitemShut
  {NoStop}%
\bibitem [{\citenamefont {Dienes}\ \emph {et~al.}(2007)\citenamefont {Dienes},
  \citenamefont {Lennek}, \citenamefont {Senechal},\ and\ \citenamefont
  {Wasnik}}]{Dienes:2007_2}%
  \BibitemOpen
  \bibfield  {author} {\bibinfo {author} {\bibfnamefont {K.~R.}\ \bibnamefont
  {Dienes}}, \bibinfo {author} {\bibfnamefont {M.}~\bibnamefont {Lennek}},
  \bibinfo {author} {\bibfnamefont {D.}~\bibnamefont {Senechal}}, \ and\
  \bibinfo {author} {\bibfnamefont {V.}~\bibnamefont {Wasnik}},\ }\href
  {\doibase 10.1103/PhysRevD.75.126005} {\bibfield  {journal} {\bibinfo
  {journal} {Phys. Rev.}\ }\textbf {\bibinfo {volume} {D75}},\ \bibinfo {pages}
  {126005} (\bibinfo {year} {2007})},\ \Eprint {http://arxiv.org/abs/0704.1320}
  {arXiv:0704.1320 [hep-th]} \BibitemShut {NoStop}%
\bibitem [{\citenamefont {Assel}\ \emph {et~al.}(2010)\citenamefont {Assel},
  \citenamefont {Christodoulides}, \citenamefont {Faraggi}, \citenamefont
  {Kounnas},\ and\ \citenamefont {Rizos}}]{Assel:2010}%
  \BibitemOpen
  \bibfield  {author} {\bibinfo {author} {\bibfnamefont {B.}~\bibnamefont
  {Assel}}, \bibinfo {author} {\bibfnamefont {K.}~\bibnamefont
  {Christodoulides}}, \bibinfo {author} {\bibfnamefont {A.~E.}\ \bibnamefont
  {Faraggi}}, \bibinfo {author} {\bibfnamefont {C.}~\bibnamefont {Kounnas}}, \
  and\ \bibinfo {author} {\bibfnamefont {J.}~\bibnamefont {Rizos}},\
  }\href@noop {} {\  (\bibinfo {year} {2010})},\ \Eprint
  {http://arxiv.org/abs/1007.2268} {arXiv:1007.2268 [hep-th]} \BibitemShut
  {NoStop}%
\bibitem [{\citenamefont {Dienes}\ and\ \citenamefont
  {Lennek}(2007)}]{Dienes:2007}%
  \BibitemOpen
  \bibfield  {author} {\bibinfo {author} {\bibfnamefont {K.~R.}\ \bibnamefont
  {Dienes}}\ and\ \bibinfo {author} {\bibfnamefont {M.}~\bibnamefont
  {Lennek}},\ }\href {\doibase 10.1103/PhysRevD.75.026008} {\bibfield
  {journal} {\bibinfo  {journal} {Phys. Rev.}\ }\textbf {\bibinfo {volume}
  {D75}},\ \bibinfo {pages} {026008} (\bibinfo {year} {2007})},\ \Eprint
  {http://arxiv.org/abs/hep-th/0610319} {arXiv:hep-th/0610319} \BibitemShut
  {NoStop}%
\bibitem [{\citenamefont {Robinson}\ \emph {et~al.}(2009)\citenamefont
  {Robinson}, \citenamefont {Cleaver},\ and\ \citenamefont
  {Hunziker}}]{Robinson:2008}%
  \BibitemOpen
  \bibfield  {author} {\bibinfo {author} {\bibfnamefont {M.}~\bibnamefont
  {Robinson}}, \bibinfo {author} {\bibfnamefont {G.}~\bibnamefont {Cleaver}}, \
  and\ \bibinfo {author} {\bibfnamefont {M.~B.}\ \bibnamefont {Hunziker}},\
  }\href {\doibase 10.1142/S0217732309031843} {\bibfield  {journal} {\bibinfo
  {journal} {Mod. Phys. Lett.}\ }\textbf {\bibinfo {volume} {A24}},\ \bibinfo
  {pages} {2703} (\bibinfo {year} {2009})},\ \Eprint
  {http://arxiv.org/abs/0809.5094} {arXiv:0809.5094 [hep-th]} \BibitemShut
  {NoStop}%
\bibitem [{\citenamefont {Obousy}\ \emph {et~al.}(2009)\citenamefont {Obousy},
  \citenamefont {Robinson},\ and\ \citenamefont {Cleaver}}]{Obousy:2008}%
  \BibitemOpen
  \bibfield  {author} {\bibinfo {author} {\bibfnamefont {R.~K.}\ \bibnamefont
  {Obousy}}, \bibinfo {author} {\bibfnamefont {M.~B.}\ \bibnamefont
  {Robinson}}, \ and\ \bibinfo {author} {\bibfnamefont {G.~B.}\ \bibnamefont
  {Cleaver}},\ }\href {\doibase 10.1142/S0217732309030965} {\bibfield
  {journal} {\bibinfo  {journal} {Mod. Phys. Lett.}\ }\textbf {\bibinfo
  {volume} {A24}},\ \bibinfo {pages} {1577} (\bibinfo {year} {2009})},\ \Eprint
  {http://arxiv.org/abs/0810.1038} {arXiv:0810.1038 [hep-ph]} \BibitemShut
  {NoStop}%
\bibitem [{\citenamefont {Kawai}\ \emph {et~al.}(1986)\citenamefont {Kawai},
  \citenamefont {Lewellen},\ and\ \citenamefont {Tye}}]{Kawai:1986}%
  \BibitemOpen
  \bibfield  {author} {\bibinfo {author} {\bibfnamefont {H.}~\bibnamefont
  {Kawai}}, \bibinfo {author} {\bibfnamefont {D.~C.}\ \bibnamefont {Lewellen}},
  \ and\ \bibinfo {author} {\bibfnamefont {S.~H.~H.}\ \bibnamefont {Tye}},\
  }\href {\doibase 10.1103/PhysRevD.34.3794} {\bibfield  {journal} {\bibinfo
  {journal} {Phys. Rev.}\ }\textbf {\bibinfo {volume} {D34}},\ \bibinfo {pages}
  {3794} (\bibinfo {year} {1986})}\BibitemShut {NoStop}%
\bibitem [{\citenamefont {Kawai}\ \emph {et~al.}(1988)\citenamefont {Kawai},
  \citenamefont {Lewellen}, \citenamefont {Schwartz},\ and\ \citenamefont
  {Tye}}]{Kawai:1987}%
  \BibitemOpen
  \bibfield  {author} {\bibinfo {author} {\bibfnamefont {H.}~\bibnamefont
  {Kawai}}, \bibinfo {author} {\bibfnamefont {D.~C.}\ \bibnamefont {Lewellen}},
  \bibinfo {author} {\bibfnamefont {J.~A.}\ \bibnamefont {Schwartz}}, \ and\
  \bibinfo {author} {\bibfnamefont {S.~H.~H.}\ \bibnamefont {Tye}},\ }\href
  {\doibase 10.1016/0550-3213(88)90544-5} {\bibfield  {journal} {\bibinfo
  {journal} {Nucl. Phys.}\ }\textbf {\bibinfo {volume} {B299}},\ \bibinfo
  {pages} {431} (\bibinfo {year} {1988})}\BibitemShut {NoStop}%
\bibitem [{\citenamefont {Lewellen}(1990)}]{Lewellen:1989}%
  \BibitemOpen
  \bibfield  {author} {\bibinfo {author} {\bibfnamefont {D.~C.}\ \bibnamefont
  {Lewellen}},\ }\href {\doibase 10.1016/0550-3213(90)90251-8} {\bibfield
  {journal} {\bibinfo  {journal} {Nucl. Phys.}\ }\textbf {\bibinfo {volume}
  {B337}},\ \bibinfo {pages} {61} (\bibinfo {year} {1990})}\BibitemShut
  {NoStop}%
\bibitem [{\citenamefont {Chaudhuri}\ \emph {et~al.}(1995)\citenamefont
  {Chaudhuri}, \citenamefont {Chung}, \citenamefont {Hockney},\ and\
  \citenamefont {Lykken}}]{Chaudhuri:1994}%
  \BibitemOpen
  \bibfield  {author} {\bibinfo {author} {\bibfnamefont {S.}~\bibnamefont
  {Chaudhuri}}, \bibinfo {author} {\bibfnamefont {S.~W.}\ \bibnamefont
  {Chung}}, \bibinfo {author} {\bibfnamefont {G.}~\bibnamefont {Hockney}}, \
  and\ \bibinfo {author} {\bibfnamefont {J.~D.}\ \bibnamefont {Lykken}},\
  }\href {\doibase 10.1016/0550-3213(95)00147-7} {\bibfield  {journal}
  {\bibinfo  {journal} {Nucl. Phys.}\ }\textbf {\bibinfo {volume} {B456}},\
  \bibinfo {pages} {89} (\bibinfo {year} {1995})},\ \Eprint
  {http://arxiv.org/abs/hep-ph/9501361} {arXiv:hep-ph/9501361} \BibitemShut
  {NoStop}%
\bibitem [{\citenamefont {Aldazabal}\ \emph {et~al.}(1995)\citenamefont
  {Aldazabal}, \citenamefont {Font}, \citenamefont {Ibanez},\ and\
  \citenamefont {Uranga}}]{Aldazabal:1994}%
  \BibitemOpen
  \bibfield  {author} {\bibinfo {author} {\bibfnamefont {G.}~\bibnamefont
  {Aldazabal}}, \bibinfo {author} {\bibfnamefont {A.}~\bibnamefont {Font}},
  \bibinfo {author} {\bibfnamefont {L.~E.}\ \bibnamefont {Ibanez}}, \ and\
  \bibinfo {author} {\bibfnamefont {A.~M.}\ \bibnamefont {Uranga}},\ }\href
  {\doibase 10.1016/0550-3213(95)00282-W} {\bibfield  {journal} {\bibinfo
  {journal} {Nucl. Phys.}\ }\textbf {\bibinfo {volume} {B452}},\ \bibinfo
  {pages} {3} (\bibinfo {year} {1995})},\ \Eprint
  {http://arxiv.org/abs/hep-th/9410206} {arXiv:hep-th/9410206} \BibitemShut
  {NoStop}%
\bibitem [{\citenamefont {Cleaver}(1995{\natexlab{a}})}]{Cleaver:1995}%
  \BibitemOpen
  \bibfield  {author} {\bibinfo {author} {\bibfnamefont {G.~B.}\ \bibnamefont
  {Cleaver}},\ }\href {\doibase 10.1016/0550-3213(95)00481-0} {\bibfield
  {journal} {\bibinfo  {journal} {Nucl. Phys.}\ }\textbf {\bibinfo {volume}
  {B456}},\ \bibinfo {pages} {219} (\bibinfo {year} {1995}{\natexlab{a}})},\
  \Eprint {http://arxiv.org/abs/hep-th/9505080} {arXiv:hep-th/9505080}
  \BibitemShut {NoStop}%
\bibitem [{\citenamefont {Cleaver}(1995{\natexlab{b}})}]{Cleaver:1995_2}%
  \BibitemOpen
  \bibfield  {author} {\bibinfo {author} {\bibfnamefont {G.~B.}\ \bibnamefont
  {Cleaver}},\ }\href@noop {} {\  (\bibinfo {year} {1995}{\natexlab{b}})},\
  \Eprint {http://arxiv.org/abs/hep-th/9506006} {arXiv:hep-th/9506006}
  \BibitemShut {NoStop}%
\bibitem [{\citenamefont {Cleaver}(1996)}]{Cleaver:1996}%
  \BibitemOpen
  \bibfield  {author} {\bibinfo {author} {\bibfnamefont {G.~B.}\ \bibnamefont
  {Cleaver}},\ }\href@noop {} {\  (\bibinfo {year} {1996})},\ \Eprint
  {http://arxiv.org/abs/hep-th/9604183} {arXiv:hep-th/9604183} \BibitemShut
  {NoStop}%
\bibitem [{\citenamefont {Kakushadze}\ and\ \citenamefont
  {Tye}(1997)}]{Kakushadze:1996}%
  \BibitemOpen
  \bibfield  {author} {\bibinfo {author} {\bibfnamefont {Z.}~\bibnamefont
  {Kakushadze}}\ and\ \bibinfo {author} {\bibfnamefont {S.~H.~H.}\ \bibnamefont
  {Tye}},\ }\href {\doibase 10.1103/PhysRevD.55.7878} {\bibfield  {journal}
  {\bibinfo  {journal} {Phys. Rev.}\ }\textbf {\bibinfo {volume} {D55}},\
  \bibinfo {pages} {7878} (\bibinfo {year} {1997})},\ \Eprint
  {http://arxiv.org/abs/hep-th/9610106} {arXiv:hep-th/9610106} \BibitemShut
  {NoStop}%
\bibitem [{\citenamefont {Erler}(1996)}]{Erler:1996}%
  \BibitemOpen
  \bibfield  {author} {\bibinfo {author} {\bibfnamefont {J.}~\bibnamefont
  {Erler}},\ }\href {\doibase 10.1016/0550-3213(96)00305-7} {\bibfield
  {journal} {\bibinfo  {journal} {Nucl. Phys.}\ }\textbf {\bibinfo {volume}
  {B475}},\ \bibinfo {pages} {597} (\bibinfo {year} {1996})},\ \Eprint
  {http://arxiv.org/abs/hep-th/9602032} {arXiv:hep-th/9602032} \BibitemShut
  {NoStop}%
\bibitem [{\citenamefont {Kakushadze}\ \emph {et~al.}(1998)\citenamefont
  {Kakushadze}, \citenamefont {Shiu}, \citenamefont {Tye},\ and\ \citenamefont
  {Vtorov-Karevsky}}]{Kakushadze:1997}%
  \BibitemOpen
  \bibfield  {author} {\bibinfo {author} {\bibfnamefont {Z.}~\bibnamefont
  {Kakushadze}}, \bibinfo {author} {\bibfnamefont {G.}~\bibnamefont {Shiu}},
  \bibinfo {author} {\bibfnamefont {S.~H.~H.}\ \bibnamefont {Tye}}, \ and\
  \bibinfo {author} {\bibfnamefont {Y.}~\bibnamefont {Vtorov-Karevsky}},\
  }\href {\doibase 10.1142/S0217751X98001323} {\bibfield  {journal} {\bibinfo
  {journal} {Int. J. Mod. Phys.}\ }\textbf {\bibinfo {volume} {A13}},\ \bibinfo
  {pages} {2551} (\bibinfo {year} {1998})},\ \Eprint
  {http://arxiv.org/abs/hep-th/9710149} {arXiv:hep-th/9710149} \BibitemShut
  {NoStop}%
\end{thebibliography}%
\end{document}